\documentclass[aps,superscriptaddress]{revtex4-2}
\usepackage{amssymb,amsmath,epsfig}
\usepackage{subfigure}
\usepackage{color}
\usepackage[colorlinks=true, pdfstartview=FitV, linkcolor=blue, citecolor=red, urlcolor=magenta, breaklinks=true]{hyperref}

\begin{document}
\title{Quasinormal modes and shadow of noncommutative black hole}

\author{J. A. V. Campos}\email{joseandrecampos@gmail.com}
\affiliation{Departamento de F\'isica, Universidade Federal da Para\'iba, 
Caixa Postal 5008, 58051-970 Jo\~ao Pessoa, Para\'iba, Brazil}
\author{M. A. Anacleto}\email{anacleto@df.ufcg.edu.br}
\affiliation{Departamento de F\'{\i}sica, Universidade Federal de Campina Grande
Caixa Postal 10071, 58429-900 Campina Grande, Para\'{\i}ba, Brazil}
\author{F. A. Brito}\email{fabrito@df.ufcg.edu.br}
\affiliation{Departamento de F\'{\i}sica, Universidade Federal de Campina Grande
Caixa Postal 10071, 58429-900 Campina Grande, Para\'{\i}ba, Brazil}
\affiliation{Departamento de F\'isica, Universidade Federal da Para\'iba, 
Caixa Postal 5008, 58051-970 Jo\~ao Pessoa, Para\'iba, Brazil}
\author{E. Passos}\email{passos@df.ufcg.edu.br}
\affiliation{Departamento de F\'{\i}sica, Universidade Federal de Campina Grande
Caixa Postal 10071, 58429-900 Campina Grande, Para\'{\i}ba, Brazil}

\begin{abstract} 
In this paper we investigate quasinormal modes (QNM) for a scalar field around a noncommutative Schwarzschild black hole. We verify  the effect of noncommutativity on quasinormal frequencies by applying two procedures widely used in the literature. The first is the Wentzel-Kramers-Brillouin (WKB) approximation up to sixth order. In the second case we use the continuous fraction method developed by Leaver. 
Besides, we also show that due to noncommutativity, the shadow radius is reduced when we increase the noncommutative parameter. In addition,  we find that the shadow radius is nonzero even at the zero mass limit for finite noncommutative parameter.
\end{abstract}

\maketitle
\pretolerance10000

\section{Introduction}

Initial studies for black hole perturbations were done by Regge and Wheeler~\cite{Regge:1957td} 
and  Zerilli~\cite{Zerilli:1970se} for Schwarzschild geometry, as well as for the Kerr black hole~\cite{Teukolsky:1972my}. In 1970 Vishveshwara~\cite{Vishveshwara:1970zz} identified a type of disturbance subject to special conditions such as outgoing waves in the spatial infinity and ingoing waves in the vicinity of the event horizon. These disturbances were called quasinormal modes valid only for a group of complex 
frequencies~\cite{Press:1971wr}. These quasinormal frequencies present a real part that provides the oscillation frequency while the imaginary part determines the damping rate of the modes. 
The dominant quasinormal modes can be seen in gravitational wave signals and 
in this case, the emitted waves are related to many physical processes such as astrophysical phenomena involving the evolution of binary systems and stellar oscillations or other highly dense objects in the early universe. 
{In this way these quasinormal modes have been observed experimentally by LIGO/VIRGO~\cite{Abbott:2016blz,TheLIGOScientific:2016src}.}

{The analysis of quasinormal modes has been widely explored in the literature~\cite{Cardoso:2003cj,Berti:2009kk,Dreyer:2002vy,Santos:2015gja,Cruz:2015bcj,Oliveira:2018oha,Cardoso:2019mqo,Moulin:2019bfh,Panotopoulos:2019gtn,Cruz:2020emz,chakraborty2018signatures} by using different mechanisms, such as the WKB approximation and numerical methods.} 
The first works using the WKB approximation to find quasinormal modes were done by Schutz and Will~\cite{Schutz:1985zz}. 
Improvements in the method were made by Iyer and Will by adding corrections up to third order~\cite{Iyer:1986np}, and so the results for Schwarzschild black hole are very close to those obtained by the numerical method of Leaver~\cite{Leaver:1985ax} in $l \geq 4$ regime.
Aiming at a new improvement in the WKB approach, studies made by Konoplya~\cite{Konoplya:2003ii} extended the method up to sixth order leading to more accurate results. Currently, we can find extensions of the WKB approximation up to  thirteenth order~\cite{Konoplya:2019hlu}.  
As aforementioned, the numerical method is another way to obtain quasinormal frequency modes.
Therefore, the first numerical approach to calculate quasinormal frequencies was described by Leaver, 
and the applied mathematical procedure is called a continuous fraction~\cite{Baber}. 
In~\cite{Leaver:1985ax}, Leaver has obtained quasinormal modes for Schwarzschild and Kerr black holes and also for Reissner-Nordstr\"{o}m black hole in~\cite{Leaver:1990zz}. 
Several works~\cite{Konoplya:2011qq,Cardoso:2004fi,Richartz:2014jla,Richartz:2015saa} have applied this numerical method that has presented a good precision.

{In addition to quasinormal modes, in recent years several authors have devoted themselves to study the shadow of the black hole~\cite{cunha2018shadows,mishra2019understanding,KONOPLYA20191,haroon2020shadow,bisnovatyi2018shadow}. This shadow requires information about the geometry around the black hole, which makes its study a very important way to understand the properties near the event horizon. Moreover with the advancement and improvements of the experimental techniques allowed us the first image of a supermassive black hole in the center of the M87 galaxy by the Event Horizon Telescope~\cite{event2019firstI,event2019firstVI} by using the properties of the shadow. These experimental results have been studied by several authors~\cite{bambi2019testing,banerjee2020silhouette,khodadi2020black} stimulated by the possibility of understanding phenomena in the regime close to the event horizon.

In this work we will use the shadow ray to better understand the proximity of the noncommutative Schwarzschild black hole horizon. 
The noncommutative gravity has been extensively investigated, particularly in black hole physics, mainly due to the possibility of better understanding the final stage of the black hole ---  see~\cite{nicolini2009noncommutative,smailagic2003feynman, smailagic2003uv, Nicolini:2005vd} for further details. 
We know that there is a relationship between quasinormal modes and of the black hole shadow and that several investigations contributed to this understanding. One of the first studies that certainly served as a basis for structuring this relationship was made by Mashhoon~\cite{mashhoon1985stability}, which describes an alternative method to calculate quasinormal modes at the eikonal limit. 
The most detailed geodesics  study is shown in Cardoso {\it et al}~\cite{cardoso2009geodesic}, which shows that the real part of the quasinormal modes is related to the angular velocity of the null circular orbit and the imaginary part is associated with the Lyapunov exponent. 
Stefanov {\it et al}~\cite{stefanov2010connection} in the eikonal regime stablished a connection between black hole quasinormal modes and lensing in the strong deflection limit. Currently, important results have been obtained at the eikonal limit, such as the relation between the real part of quasinormal frequencies and the black hole shadow radius~\cite{jusufi2020quasinormal,cuadros2020analytical,Moura:2021eln}.}

Studies related to quasinormal modes of noncommutative black holes have been extensively carried out by several authors~\cite{Giri:2006rc,Gupta:2017lwk,Gupta:2015uga,Liang:2018uyk,Liang:2018nmr,Ciric:2017rnf}. 
In this paper, we aim to determine the quasinormal modes of noncommutative Schwarzschild black hole via Lorentzian mass distribution in order to verify the changes caused by the noncommutative parameter. 
{Moreover, we show that contrary to the case of the Schwarzschild black hole, the shadow radius presents 
a non-zero result at the zero mass limit. 
 Therefore, at this limit, the shadow radius is proportional to a minimum mass.
{This result has not been obtained analytically by using the Gaussian distribution. 
Thus, considering the Lorentzian distribution, some results in an analytical way are more easily explored than in the Gaussian case where this is done numerically. }
Furthermore, in~\cite{Anacleto:2020zhp,Anacleto:2020lel}, we have also found a similar result when investigating the zero mass limit in the black hole absorption process. In that case, we have obtained a non-zero absorption at the zero mass limit.
By considering a Lorentzian mass distribution to introduce the noncommutativity, 
in~\cite{Anacleto:2019tdj}, we have explored the process of scattering and absorption of scalar waves through 
a noncommutative Schwarzschild black hole. Moreover, 
in~\cite{Anacleto:2020efy,Anacleto:2020zfh,Anacleto:2015kca,Anacleto:2014cga,Nozari:2009nr,Mehdipour:2009zz,Mehdipour:2010kp,Mehdipour:2010ap,Miao:2010wy,Miao:2011dy,Nozari:2012bp,Ovgun:2015box,Gecim:2020zcb,Rahaman:2013gw,Sadeghi:2015nzp,Liang:2012vx}, the thermodynamics of the BTZ and Schwarzschild black holes in the noncommutative background has been investigated by using the WKB approach in tunneling formalism~\cite{Anacleto:2014apa,Anacleto:2015awa,Anacleto:2015mma}. }
{An advantage of using the Lorentzian distribution in analytical calculus has been investigated in ~\cite{Anacleto:2020efy} 
and~\cite{Anacleto:2020zfh} where logarithmic corrections for entropy 
and the condition for black hole remnant formation were obtained.

We organize the paper as follows:
In Sec.~\ref{sec2} we implemented the effect of noncommutativity in the Schwarzschild black hole metric by a Lorentzian smeared mass distribution, and we analyze the results for quasinormal frequencies.
In Sec.~\ref{sec3} we apply the null geodetic method to determine the shadow of the noncommutative black hole.
In Sec.~\ref{conc} we make our final considerations.

\section{Noncommutative black hole with Lorentzian smeared mass distribution}\label{sec2}
On this section we begin by considering a Lorentzian distribution \cite{Nicolini:2005vd,Nozari:2008rc} given by
\begin{eqnarray}
\rho_{\theta}(r)=\frac{M\sqrt{\theta}}{\pi^{3/2}(r^2+\pi\theta)^{2}},
\end{eqnarray}
where $ \theta $ is the noncommutative parameter of dimension ${length}^2$ and $ M $ is the total mass diffused throughout the region
of linear size $ \sqrt{\theta} $. 
Thus, the smeared mass distribution function becomes~\cite{Anacleto:2019tdj} 
\begin{eqnarray}
{\cal M}_{\theta}&=&\int_0^r\rho_{\theta}(r)4\pi r^2 dr,
\\
&=&\frac{2M}{\pi}\left[\tan^{-1}\left( \frac{r}{\sqrt{\pi\theta}} \right)
-\frac{r\sqrt{\pi\theta}}{\pi\theta + r^2}  \right],
\\
&=&M-\frac{4 M\sqrt{\theta}}{\sqrt{\pi}r} + {\cal O}(\theta^{3/2}). 
\end{eqnarray}
Hence, the line element of the Schwarzschild black hole in the noncommutative background is now given by
\begin{eqnarray}
ds^{2} = -f(r)dt^{2} + f(r)^{-1}dr^{2}+r^{2}d\Omega^2,
\end{eqnarray}
with
\begin{eqnarray}
f(r) &=& 1 - \frac{2 M}{r} + \frac{8M\sqrt{\theta}}{\sqrt{\pi}r^{2}},
\\
&=&\frac{1}{r^{2}}\left(r - r_{+}\right)\left(r - r_{-}\right),
\label{elbg}
\end{eqnarray}
where 
\begin{eqnarray}
&&r_{+} = M + \sqrt{M^{2} - 8M\sqrt{\theta/\pi}}\approx 2M - 4 \sqrt{\frac{\theta}{\pi}},
\\
&&r_{-} = M - \sqrt{M^{2} - 8M\sqrt{\theta/\pi}}\approx 4 \sqrt{\frac{\theta}{\pi}},
\end{eqnarray}
which represent the radius of the event horizon and the Cauchy horizon, respectively.

The next step we consider the case of the massless scalar field described by the  Klein-Gordon equation in the background \eqref{elbg}
\begin{equation}
\frac{1}{\sqrt{-g}}\partial_{\mu}\left(\sqrt{-g}g^{\mu\nu}\partial_{\nu}\right)\Psi = 0.
\end{equation}
Thus, we apply the separation of variables  method in the above equation by using the following Ansatz
\begin{eqnarray}
\Psi(\textbf{r},t) = \dfrac{R_{\omega l}(r)}{r}Y_{lm}(\vartheta, \phi)e^{-i\omega t},
\end{eqnarray}
where $\omega$ is the frequency and $Y_{lm}(\vartheta, \phi)$ are the spherical harmonics.

Now, we can obtain a radial equation for $R_{\omega l}(r)$:
\begin{equation}
 \left(r - r_{-}\right)\left(r - r_{+}\right)\dfrac{d^{2}R_{\omega l}(r)}{dr^{2}} + r^{2} \dfrac{df(r)}{dr}\dfrac{dR_{\omega l}(r)}{dr} + \left[\dfrac{\omega^{2}r^{4}}{\left(r - r_{-}\right)\left(r - r_{+}\right)} - r \dfrac{df(r)}{dr} - l(l + 1)\right]R_{\omega l}(r) = 0.
\label{ER}
\end{equation} 

We can reduce the radial equation \eqref{ER} into a Schr\"{o}dinger like equation by introducing a new coordinate (called tortoise coordinate) given by  $dr_{*} = f(r)^{-1}dr$ and
\begin{equation}
r_{*}= r + \dfrac{r_{-}^{2}}{r_{-} - r_{+}}\log(r - r_{-}) - \dfrac{r_{+}^{2}}{r_{-} - r_{+}}\log(r - r_{+}),
\end{equation}
so that the radial equation becomes
\begin{eqnarray}
\dfrac{d^{2}R_{\omega l}(r_{*})}{dr_{*}^{2}} + \left[\omega^{2} - V_{eff}\right]R_{\omega l}(r_{*})=0,
\label{ERtor}
\end{eqnarray}
where 
\begin{equation}
V_{eff}=\dfrac{f(r)}{r}\dfrac{df(r)}{dr} + \dfrac{f(r)l(l+1)}{r^{2}}.
\end{equation}
In the following sections, we will obtain the quasinormal frequencies by two methods that are widely used in the literature. The first uses a sixth order WKB approximation, and the second method introduced by Leaver and improved by Nollert~\cite{Nollert:1993zz},  consists of using the continuous fraction method to find numerically quasinormal modes.

\subsection{ WKB approximation}
Quasinormal modes correspond to solutions of the wave equation \eqref{ERtor} that satisfy the conditions of the purely outgoing waves at infinity and purely incoming waves at the event horizon, i.e.,
\begin{equation}
R_{\omega l}(r_{*}) \sim e^{\pm i\omega r_{*}}, \qquad r_{*}\rightarrow \pm\infty.
\end{equation}
In this section we will use the WKB approximation to find the quasinormal modes. The first works using the WKB approximation to evaluate the quasinormal modes were done by Schutz and Will~\cite{Schutz:1985zz}. Improvements in the method were made using corrections up to 
third order~\cite{Iyer:1986np,Seidel:1989bp} and up to sixth order by Konoplya~\cite{Konoplya:2003ii}.
The quasinormal modes are obtained by using the sixth order corrections for the WKB approximation as follows 
\begin{equation}
\dfrac{i\left(\omega_{n}^{2} - \bar{V}_{eff}\right)}{\sqrt{-2\bar{V}_{eff}''}} - \sum_{j=2}^{6}\Omega_{j}= n + \dfrac{1}{2},
\label{eqWKB}
\end{equation}
where $\Omega_{j}$ are the correction terms of the model. We have that $\bar{V}_{eff}$ is the maximum effective potential at point $\bar{r}_{*}$ and $('')$ refers to the second derivative with respect to the tortoise coordinate. We can obtain the values of $\bar {r}_{*}$ by making $ \bar{V}_{eff}' = 0$. In figure \ref{poten} we show the curves of the effective potential for $l = 1, 2$ and $\Theta = 0.0, 0.05, 0.10, 0.12 $, where we define $\Theta = \sqrt{\theta}/(M\sqrt{\pi})$ for $M = 1$. In the tables (\ref{tab1} - \ref{tab3}) we present the tabulated quasinormal frequencies using sixth order WKB method. 
\begin{figure}[!htb]
 \centering
\subfigure[]{\includegraphics[scale=0.45]{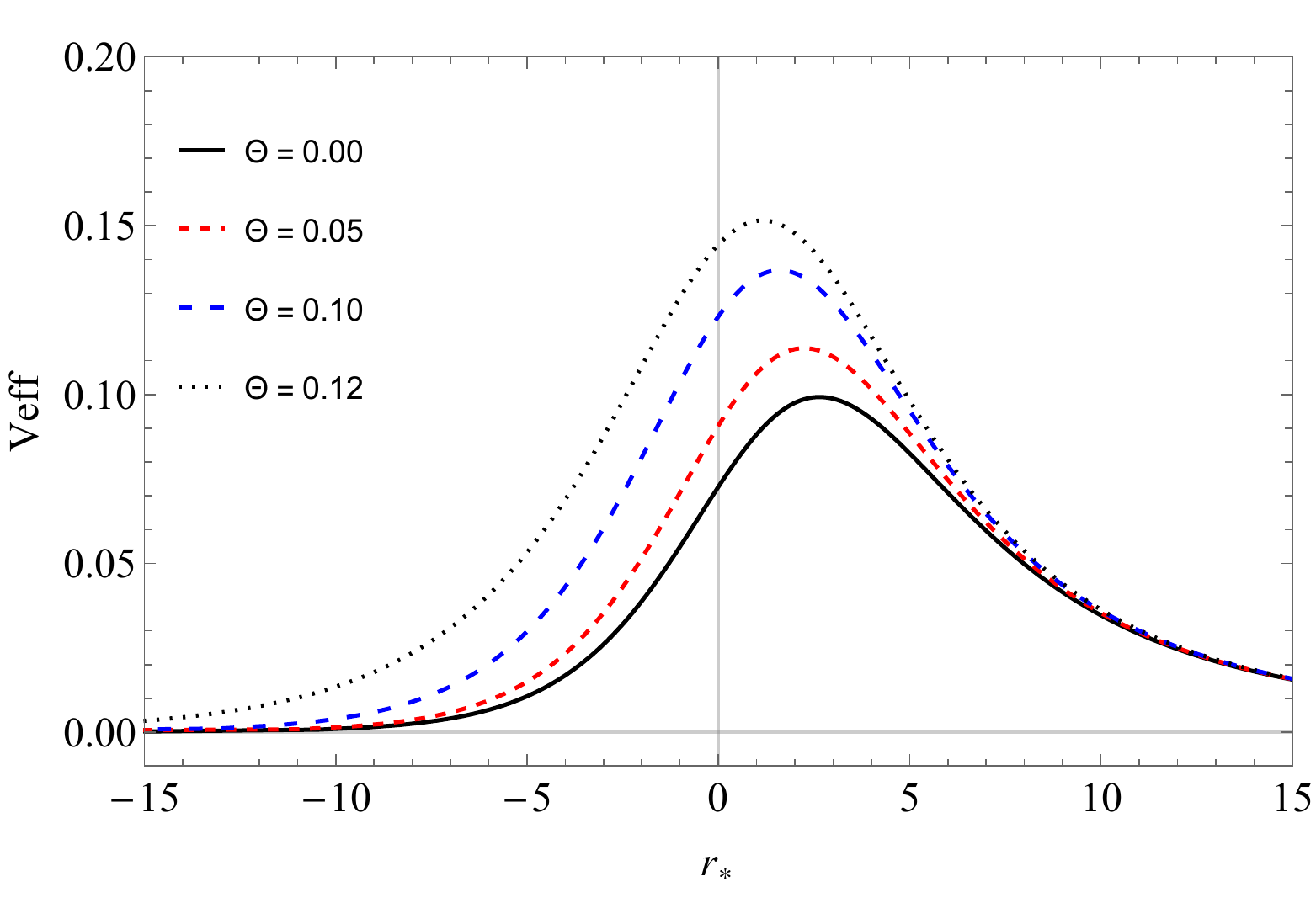}\label{fig1}}
\qquad
\subfigure[]{\includegraphics[scale=0.45]{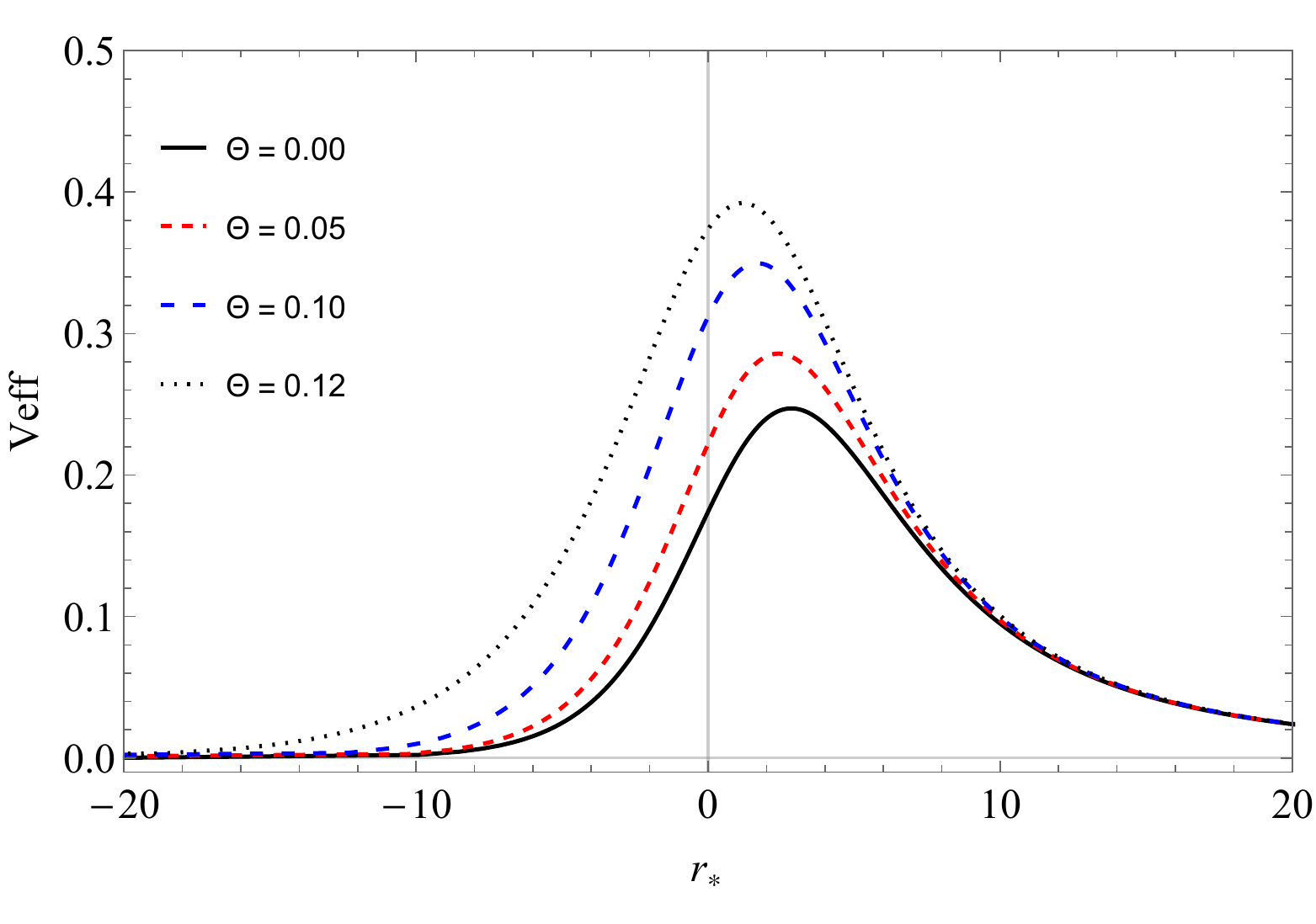}\label{fig2}}
 \caption{\footnotesize{The effective potential $V_{eff}$ as function of the tortoise coordinate  $r_{*}$ (a) $l = 1$ and (b) $l = 2$. }} 
 \label{poten}
\end{figure}
Another possibility is to study scattering by using the WKB method done in~\cite{Iyer:1986np}.  In order to develop this investigation, we use the boundary conditions for equation \eqref{ERtor} in the form:
\begin{equation}
R_{\omega l}(r_{*}) = \begin{cases}
A_{int}e^{-i\omega r_{*}} + A_{out}e^{i\omega r_{*}}, \qquad r_{*} \rightarrow \infty ,\\
A_{tr}e^{-i\omega r_{*}}, \hspace{2.7cm} r_{*} \rightarrow -\infty.
\end{cases}
\label{condtor}
\end{equation}
Notice that as we want to obtain the reflection and transmission coefficients, as done for tunneling in quantum mechanics, we need the condition $A_{int} \neq 0$. The quantity $\left(\omega^{2} - V_{eff}\right)$ in $\eqref{ERtor}$ is assumed to be purely real, and with these imposed conditions we can find
\begin{equation}
\mathcal{K} = \dfrac{i\left(\omega^{2} - \bar{V}_{eff}\right)}{\sqrt{-2\bar{V}_{eff}''}}  - \sum_{j=2}^{6}\Omega_{j}(\mathcal{K}),
\label{eqtr}
\end{equation}
where $\omega$ is purely real and $ \Omega_{j}(\mathcal{K})$ are coefficients that depend on the effective potential and $ \mathcal{K} $ is a purely imaginary quantity. This way of studying scattering by using the WKB approximation can be also found in~\cite{Konoplya:2019hlu,Konoplya:2019xmn}. Thus, using the relationship between $\mathcal{K}$ and the reflection and transmission coefficients obtained in~\cite{Iyer:1986np} we get the following:
\begin{eqnarray}
|R|^{2} = \dfrac{|A_{out}|^{2}}{|A_{int}|^{2}} &=& \dfrac{1}{1 + e^{-2 i \pi \mathcal{K}}}, \qquad 0 < |R|^{2} < 1,
\\
|T|^{2} = \dfrac{|A_{tr}|^{2}}{|A_{int}|^{2}}  &=& \dfrac{1}{1 + e^{2 i \pi \mathcal{K}}} = 1 - |R|^{2}.
\end{eqnarray}
Now to find the coefficients, we just calculate the value $ \mathcal{K} $ that can be obtained by solving the equation \eqref{eqtr}. This method has a good approximation for $ l > 0 $ as we can see in Figure \ref{Trs}, where we have a comparison between the numerical results and the WKB approximation for the transmission coefficient for $ l = 1,2,3 $ and $ \Theta = 0.05, 0.12 $. 
Notice that the curves obtained by the WKB approximation are very close to those obtained by the numerical method used in the paper~\cite{Anacleto:2019tdj}, showing that the method presents excellent results. The sixth-order WKB approximation does not show good results for $ l= 0$, improving only when we take large $ \omega $. This problem is also mentioned in~\cite{Konoplya:2019hlu}.
\begin{figure}[!htb]
 \centering
\subfigure[]{\includegraphics[scale=0.45]{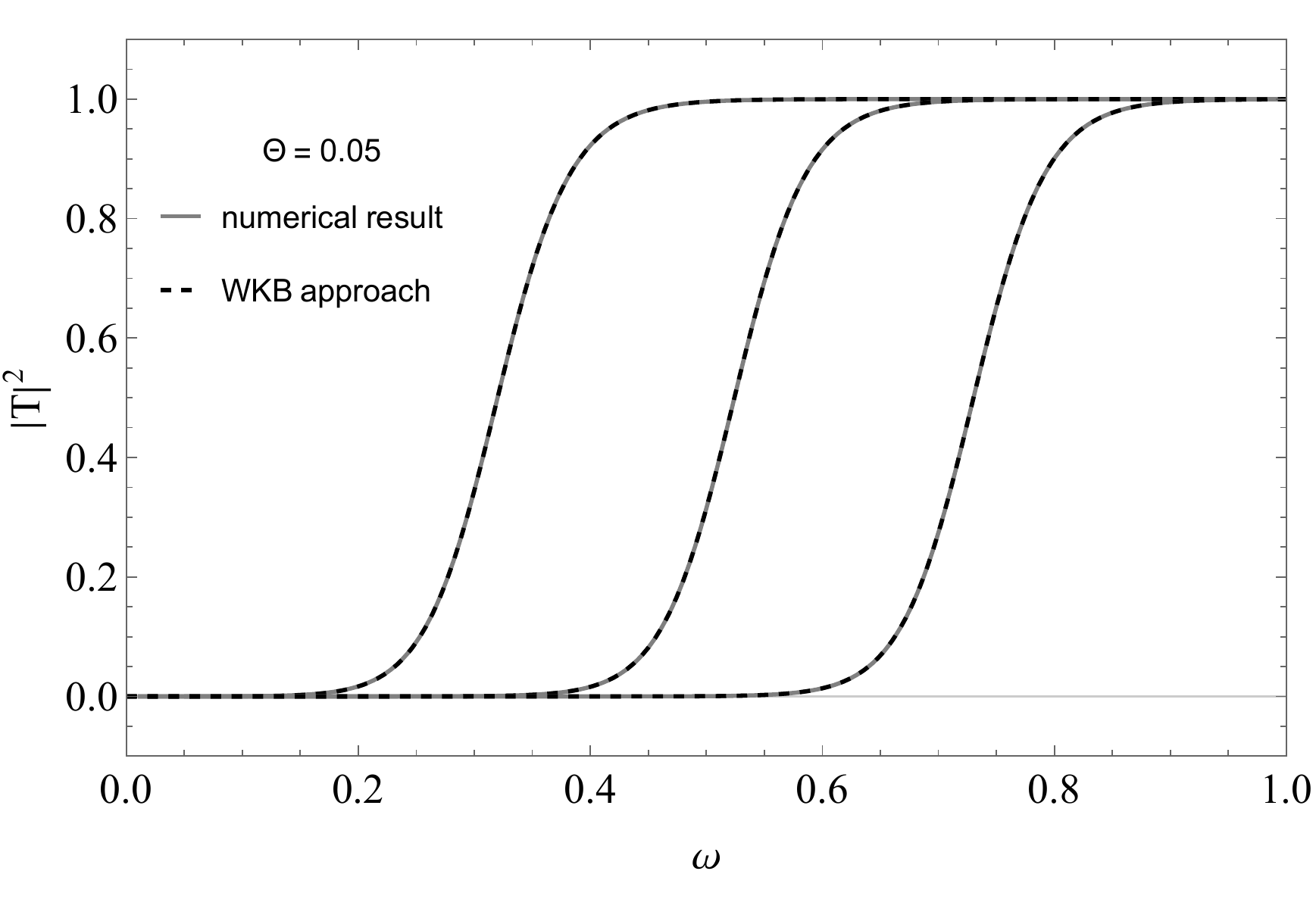}\label{T1}}
\qquad
\subfigure[]{\includegraphics[scale=0.45]{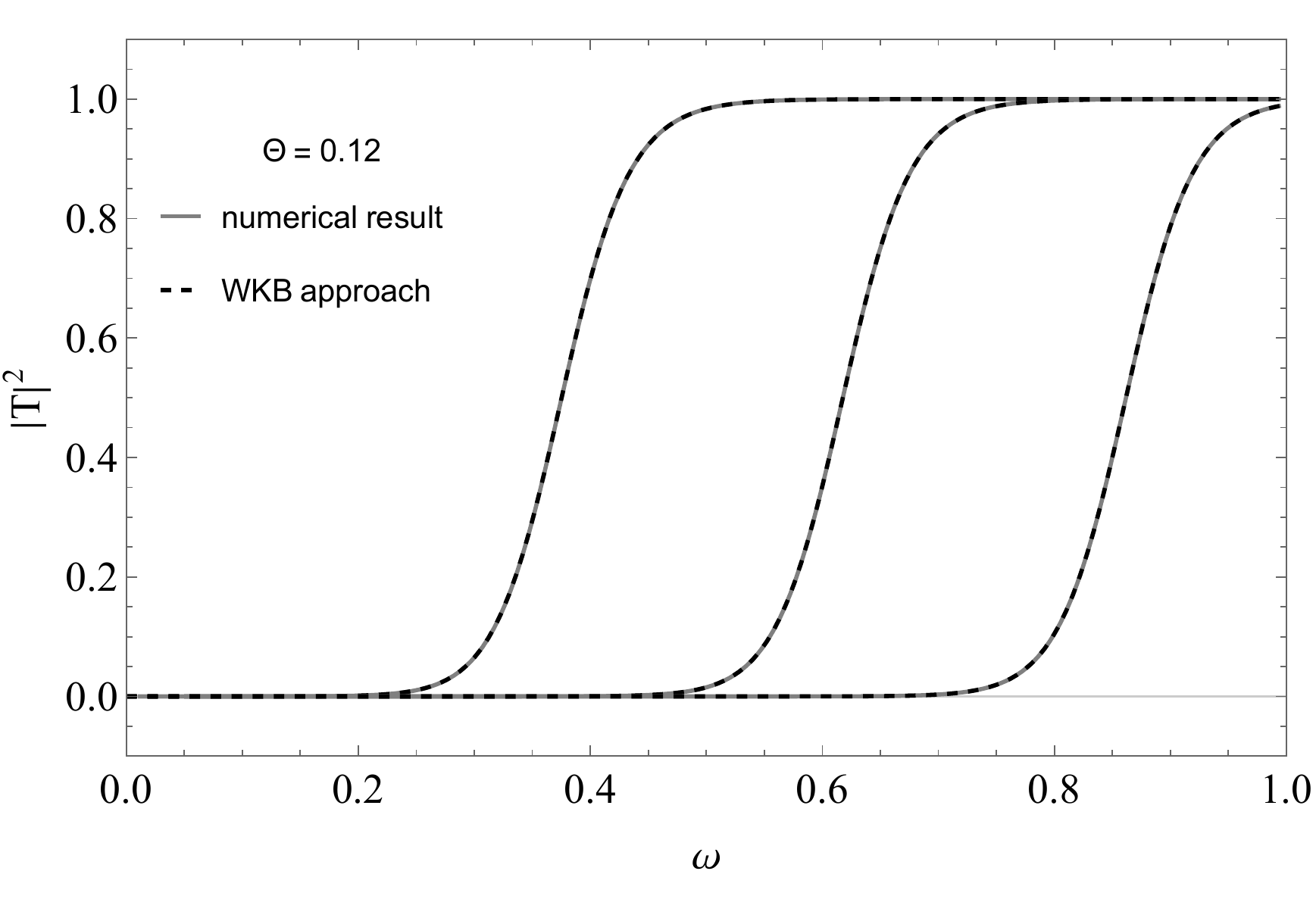}\label{T2}}
 \caption{\footnotesize{Transmission coefficients for three multipole $l=1,2,3$ (from left to right) and $\Theta = 0.05, 0.12$. The approximation between the two methods is very good.}} 
 \label{Trs}
\end{figure}

\subsection{Leaver's continued fraction}
The numerical method is another way to obtain the quasinormal frequency, and this procedure has been described by Leaver~\cite{Leaver:1985ax,Leaver:1990zz}, and which is also found in other works~\cite{Konoplya:2011qq,Cardoso:2004fi} showing that it is a method with good precision.

We start analyzing the radial equation \eqref{ER} which is subject to boundary conditions at infinity $r \rightarrow \infty$ and near the event horizon $r \rightarrow r_{+}$, such that one obtains the asymptotic solutions
\begin{equation}
R_{\omega l}(r) \approx \begin{cases}
e^{i\omega r}r^{i\omega(r_{-} +r_{+})}, \hspace{6.45cm} r \rightarrow \infty , \\
e^{-i\omega r_{+}}\left(r_{+} - r_{-}\right)^{-i\omega r_{-}^{2}/(r_{-} - r_{+})}\left(r - r_{+}\right)^{i\omega r_{+}^{2}/(r_{-} - r_{+})}, \qquad r \rightarrow r_{+}.
\end{cases}
\end{equation}
We can obtain a solution that has the desired behavior on the horizon
($ r = r _ {+} $), and  that can be written in the form
\begin{equation}
R = \frac{r}{r - r_{-}}\left(r - r_{-}\right)^{i \omega(r_{+} + r_{-})}\left(\frac{r - r_{+}}{r - r_{-}}\right)^{\frac{-i \omega r_{+}^{2}}{r_{+} - r_{-}}} e^{i\omega r}\sum_{k=0}^{\infty}a_{k}\left(\frac{r - r_{+}}{r - r_{-}}\right)^{k}.
\label{anzart}
\end{equation}
By replacing the solution \eqref{anzart} in the equation \eqref{ER}, we obtain the recurrence relation 
\begin{equation}
\alpha_{0}a_{1} + \beta_{0}a_{0} = 0,
\label{cond1}
\end{equation}
and
\begin{equation}
\alpha_{k}a_{k+1} + \beta_{k}a_{k} + \gamma_{k}a_{k-1} = 0, \qquad k \geq 1.
\label{cond2}
\end{equation}
The recurrence relation coefficients $\alpha_{k}$, $\beta_{k}$ and $\gamma_{k}$ are simple functions of $k$ and the parameters $\omega$, $l$ and the radius $r_{-}$ and $r_{+}$: 
\begin{eqnarray}
\alpha_{k} &=& \left(1 + k\right)\left[r_{-}\left(k + 1\right) - r_{+}\left(1 + k - 2i\omega r_{+}\right)\right], 
\label{cofrel1}\\
\beta_{k} &=& r_{+}\left[l(l+1) + 1 + 2\left(k - 2i\omega r_{+}\right)\left(1 + k - 2i\omega r_{+}\right)\right] - r_{-}\left[l(l + 1) + 1 - 2i\omega r_{+} + 2k\left(1 + k - 2i\omega r_{+}\right)\right],
\label{cofrel2}\\
\gamma_{k} &=& \left[k -2i\omega\left(r_{-} + r_{+}\right)\right]\left[k\left(r_{-} - r_{+}\right) + 2i\omega r_{+}^{2}\right].
\label{cofrel3}
\end{eqnarray}
{See that the solution \eqref{anzart} applied directly to the radial equation returns a three-term recurrence relation which makes it easier to use continued fractions. This is because our metric for a non-commutative black hole results in a generalized spheroidal wave equation whose solutions are connected by three-term recurrence relations --- see Appendix~\ref{Append} for further details. However, the Leaver's method is not necessarily limited to recurrence relations of this type as shown in~\cite{Leaver:1990zz}, where it was considered an equation that describes odd-parity perturbations of a charged black hole, which equation has series solutions whose coefficients are connected by four-term recurrence relations.}

The boundary condition at infinity will be satisfied for quasinormal frequency values $\omega = \omega_{n}$, so that the series in \eqref{anzart} is absolutely convergent. Hence, we have a recurrence relation with three terms to determine the coefficient $a_{k}$, and we can write in terms of a continuous fraction~\cite{Gautschi}
\begin{equation}
\dfrac{a_{k+1}}{a_{k}}=\dfrac{-\gamma_{k+1}}{\beta_{k+1} - \dfrac{\alpha_{k+1}\gamma_{k+2}}{\beta_{k+2}-\dfrac{\alpha_{k+2}\gamma_{k+3}}{\beta_{k+2} - \cdots}}},
\end{equation}
which can also be found as follows
\begin{equation}
\dfrac{a_{k+1}}{a_{k}} = \dfrac{-\gamma_{k+1}}{\beta_{k+1}-}\dfrac{\alpha_{k+1}\gamma_{k+2}}{\beta_{k+2} -}\dfrac{\alpha_{k+2}\gamma_{k+3}}{\beta_{k+3} -\cdots} .
\label{efracont}
\end{equation}
We can obtain the characteristic equation for quasinormal frequencies by assigning $k=0$ in~\eqref{efracont} and comparing with reason $a_{1}/a_{0} = -\beta_{0}/\alpha_{0}$ obtained from \eqref{cond1},
\begin{equation}
0 = \dfrac{\beta_{0}}{\alpha_{0}} - \dfrac{\gamma_{1}}{\beta_{1}-}\dfrac{\alpha_{1}\gamma_{2}}{\beta_{2} -}\dfrac{\alpha_{2}\gamma_{3}}{\beta_{3} - \cdots}.
\label{efracontcarac}
\end{equation}
With the equation above, we can obtain the quasinormal frequencies $\omega_{n}$, by just calculating its roots numerically. However, the  equation \eqref{efracontcarac} is more used to find the fundamental frequency in the case of the more stable root, and another way of finding the modes is to invert this equation to a large number of $k$ as follows
\begin{equation}
\beta_{k} - \dfrac{\alpha_{k-1}\gamma_{k}}{\beta_{k-1}-}\dfrac{\alpha_{k-2}\gamma_{k-1}}{\beta_{k-2} -}\cdots \dfrac{\alpha_{0}\gamma_{1}}{\beta_{0}} = \dfrac{\alpha_{k}\gamma_{k+1}}{\beta_{k+1}-}\dfrac{\alpha_{k+1}\gamma_{k+2}}{\beta_{k+2} -} \cdots .
\end{equation}
To complement the analysis, we will check the behavior for very large $k$, as done in~\cite{Nollert:1993zz}. 
Now, we have to reorganize the equation \eqref{cond2} dividing $\alpha$, $\beta$ and $\gamma$ by $a_k$ to obtain
\begin{equation}
\alpha_{k}\dfrac{a_{k+1}}{a_{k}} + \beta_{k} + \gamma_{k}\dfrac{a_{k-1}}{a_{k}} = 0. 
\label{cond3}
\end{equation}
We can see that, $\lim\limits_{k \to \infty}(a_{k + 1}/a_{k})\simeq 1$, and we can get a more complete expression by expanding $a_{k + 1}/a_{k}$ in power series in terms of $\sqrt{k}$,
\begin{equation}
\lim\limits_{k \to \infty}\dfrac{a_{k + 1}}{a_{k}} = \sum_{i=0}^{\infty} C_{i}k^{-i/2} = C_{0} + \dfrac{C_{1}}{\sqrt{k}} + \dfrac{C_{2}}{k} + \dfrac{C_{3}}{k^{3/2}} + \cdots .
\end{equation} 
By considering the series up to the third term and admitting $C_{0} = 1$ we have
\begin{eqnarray}
\lim\limits_{k \to \infty}\dfrac{a_{k + 1}}{a_{k}} & \approx & 1 + \dfrac{C_{1}}{\sqrt{k}} +\dfrac{C_{2}}{k} 
+\cdots ,
\\
\lim\limits_{k \to \infty}\dfrac{a_{k - 1}}{a_{k}} & \approx & 1 - \dfrac{C_{1}}{\sqrt{k}} +\dfrac{(C_{1})^{2} - C_{2}}{k} + \dfrac{C_{1}\left(2C_{2} - 1/2 - (C_{1})^{2}\right)}{k^{3/2}}+ \cdots .
\end{eqnarray}
Now, to obtain the values of $C_{1}$ and $C_{2}$, we replace these two expressions above into equation \eqref{cond3} in the very large $k$ regime by making the multiplications and restricting up to terms of the order $k^{-3/2}$ to get
\begin{equation}
\dfrac{(C_{1})^{2} - 2i\omega \left(r_{-} - r_{+}\right)}{k} - \dfrac{C_{1}}{2}\dfrac{\left[-3 + 2(C_{1})^{2} - 4C_{2} - 4i\omega \left(r_{-} + r_{+}\right)\right]}{k^{3/2}} \approx 0, 
\label{cond4}
\end{equation}
such that we find the following:
\begin{eqnarray}
\left(C_{1}\right)^{2}&=& 2i\omega \left(r_{-} - r_{+} \right),
\\
 C_{2}  &=& -2i\omega r_{+} - \dfrac{3}{4}.
 \end{eqnarray}
Now we have,
\begin{equation}
 \lim\limits_{k \to \infty}\dfrac{a_{k+1}}{a_{k}} \approx 1 \pm \dfrac{\sqrt{2i\omega \left(r_{-} - r_{+}\right)}}{\sqrt{k}} - \dfrac{2i\omega r_{+} +3/4}{k}.
\end{equation}
Here, we see that at the limit $r_{-} \rightarrow 0$ and $r_{+}\rightarrow 1$, we get the result for the Schwarzschild case initially found by Leaver.
\subsection{Results}
In the Tables (\ref{tab1} - \ref{tab3}), we present some results for the quasinormal modes computed using the sixth-order WKB approximation, and the Leaver's continues fraction method described in the previous section by admitting $M = 1$ for various values of $ \Theta = \sqrt {\theta/\pi} $, $l$ and $n$. 
We see that the results between the methods approach when $l > 1$, this is, due to the instability of the WKB method for small multipole numbers mainly close to zero this instability can also be seen 
in the graphs of Fig.~\ref{fig_Qn}. 
An important detail is in the sign of the imaginary part that is always negative when the frequency is associated with the scalar field. A justification for this is due to the exponential drop of the quasinormal modes over time by losing energy in the form of scalar waves.

We can see the influence in the quasinormal modes for the noncommutative case by admitting values for $\Theta$ where $\Theta=0$ returns to the Schwarzschild case. With the increase of the non-commutative parameter $\Theta$, we have an increase in the real part of the quasinormal frequency, while the imaginary part begins to grow and then decreases.
Another way of visualizing the effects of the noncommutative parameter is through the graphs of Fig.~\ref{fig_Qn}, these plots were obtained using the WKB method. We have quasinormal modes where in the plot we depicted the real part (top) and the imaginary part (bottom). The modes are based on $n$ for the following multipoles numbers $l = 1, 2, 3 $ and $ 4 $, for which  
we see that the results become more linear with the results varying $\Theta$  and $l = 3$ and $4$.
{Thus, in Fig.~\ref{fig_Qn} we observe that, by varying $l$ and the parameter $\theta$, the imaginary part of the frequency does not cross the horizontal axis or change sign, thus indicating that the black hole remains stable due to scalar perturbation.}
In addition, it is interesting to make a plot for the complex plane as in Fig.~\ref{plaxoplx} where we consider three families of multipoles $l = 1, 2, 3$ and varying $\Theta$ as follows $0$ (black), $0.05$ (red), $0.10$ (blue), $0.12$ (green). The left panel was obtained by the WKB approximation and the right panel by the continuous fraction method (numerical). We can see that the frequency curves incline more closely when we use the WKB approximation. We also see that for the extreme case $\Theta = 0.12$ the curve tilts more to the left in both methods.
\begin{table}[h!]
\begin{center}
 \begin{footnotesize}
\caption{\footnotesize{QN frequencies for $l = 1$ }} 
\label{tab1}
\begin{tabular}{c||c|c|c|c|c|c}
\hline
\multicolumn{1} {c||}{}&\multicolumn{2} {|c|}{ $\omega_{0}$ } & \multicolumn{2} {|c|}{ $\omega_{1}$ } & \multicolumn{2} {|c}{ $\omega_{2}$ } \\
\hline
 $\Theta$     & 6th order WKB &  numerical & 6th order WKB & numerical & 6th order WKB & numerical\\
 \hline
 0.00  & 0.292910 - 0.097762i & 0.292936 - 0.0976600i & 0.264471 - 0.306518i & 0.264449 - 0.306257i & 0.231014 - 0.542166i & 0.229539 - 0.540133i  \\
 0.05  & 0.316239 - 0.099311i & 0.316243 - 0.0992441i & 0.290671 - 0.309586i & 0.290591 - 0.309431i & 0.260297 - 0.543374i & 0.258427 - 0.542162i \\
 0.10  & 0.351481 - 0.097397i & 0.351435 - 0.0973776i & 0.330103 - 0.300074i & 0.329892 - 0.300209i & 0.300773 - 0.519039i & 0.299190 - 0.519345i \\
 0.12  & 0.371967 - 0.091934i & 0.371932 - 0.0919646i & 0.346286 - 0.282086i & 0.346358 - 0.282153i & 0.300083 - 0.491675i & 0.301082 - 0.491385i \\
 \hline
 \end{tabular}
 \end{footnotesize}
\end{center}
\end{table}
\begin{table}[h!]
\begin{center}
\begin{footnotesize}
\caption{\footnotesize{QN frequencies for $l = 2$ }}
\label{tab2}
\begin{tabular}{ c||c|c|c|c|c|c  }
 \hline
\multicolumn{1} {c||}{}&\multicolumn{2} {|c|}{ $\omega_{0}$ } & \multicolumn{2} {|c|}{ $\omega_{1}$ } & \multicolumn{2} {|c}{ $\omega_{2}$ } \\
\hline
$\Theta$     & 6th order WKB &  numerical & 6th order WKB & numerical & 6th order WKB & numerical\\
 \hline
 0.00  & 0.483642 - 0.096766i  & 0.483644 - 0.0967588i & 0.463847 - 0.295627i & 0.463851 - 0.295604i & 0.430386 - 0.508700i & 0.430544 - 0.508558i\\
 0.05  & 0.521842 - 0.098436i  & 0.521844 - 0.0984288i & 0.504088 - 0.300022i & 0.504087 - 0.299997i & 0.474038 - 0.514099i & 0.474067 - 0.513965i\\
 0.10  & 0.580028 - 0.096808i  & 0.580028 - 0.0968027i & 0.565520 - 0.293676i & 0.565511 - 0.293660i & 0.540100 - 0.499152i & 0.540039 - 0.499083i\\
 0.12  & 0.615658 - 0.091534i  & 0.615659 - 0.0915307i & 0.599809 - 0.276911i & 0.599817 - 0.276896i & 0.569259 - 0.469194i & 0.569307 - 0.469133i\\
 \hline
 \end{tabular}
 \end{footnotesize}
\end{center}
\end{table}
\begin{table}[h!]
\begin{center}
\begin{footnotesize}
\caption{\footnotesize{QN frequencies for $l = 3$ }}
\label{tab3}
\begin{tabular}{ c||c|c|c|c|c|c  }
 \hline
\multicolumn{1} {c||}{}&\multicolumn{2} {|c|}{ $\omega_{0}$ } & \multicolumn{2} {|c|}{ $\omega_{1}$ } & \multicolumn{2} {|c}{ $\omega_{2}$ } \\
\hline
$\Theta$     & 6th order WKB & numerical & 6th order WKB & numerical & 6th order WKB & numerical\\
 \hline
 0.00  & 0.675366 - 0.096501i  & 0.675366 - 0.0964996i & 0.660671 - 0.292288i & 0.660671 - 0.292285i & 0.633591 - 0.496011i & 0.633626 - 0.496008i\\
 0.05  & 0.728594 - 0.098197i  & 0.728594 - 0.0981957i & 0.715414 - 0.297060i & 0.715414 - 0.297056i & 0.691124 - 0.502924i & 0.691137 - 0.502911i\\
 0.10  & 0.809869 - 0.096642i  & 0.809869 - 0.0966413i & 0.799161 - 0.291636i & 0.799159 - 0.291633i & 0.779124 - 0.491455i & 0.779117 - 0.491442i\\
 0.12  & 0.860294 - 0.091410i  & 0.860294 - 0.0914089i & 0.848871 - 0.275415i & 0.848871 - 0.275412i & 0.826463 - 0.462957i & 0.826463 - 0.462950i\\
 \hline
 \end{tabular}
 \end{footnotesize}
\end{center}
\end{table}

\begin{figure}[!htb]
 \centering
 \subfigure[]{\includegraphics[scale=0.35]{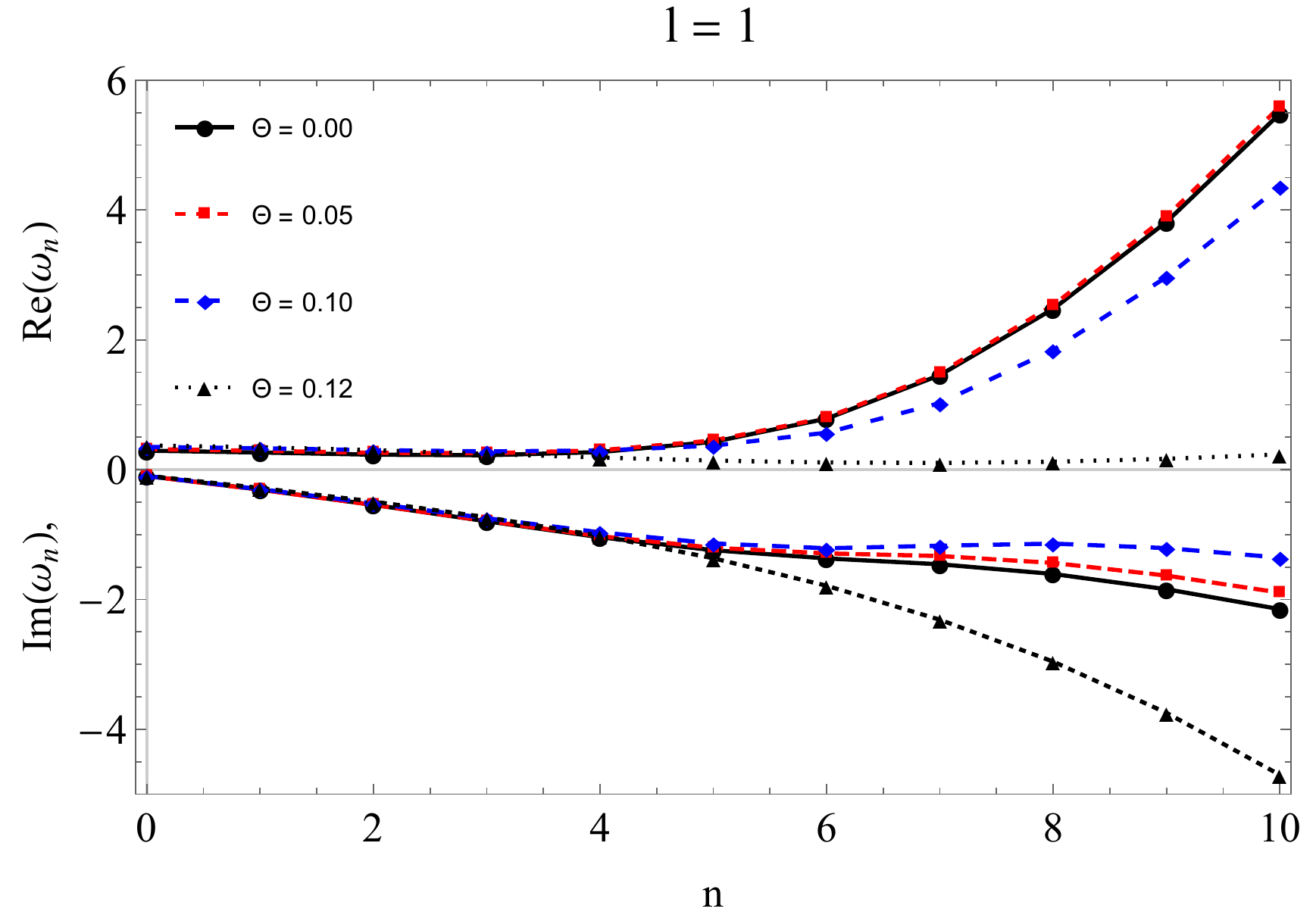}\label{figQN1}}
 \qquad
 \subfigure[]{\includegraphics[scale=0.35]{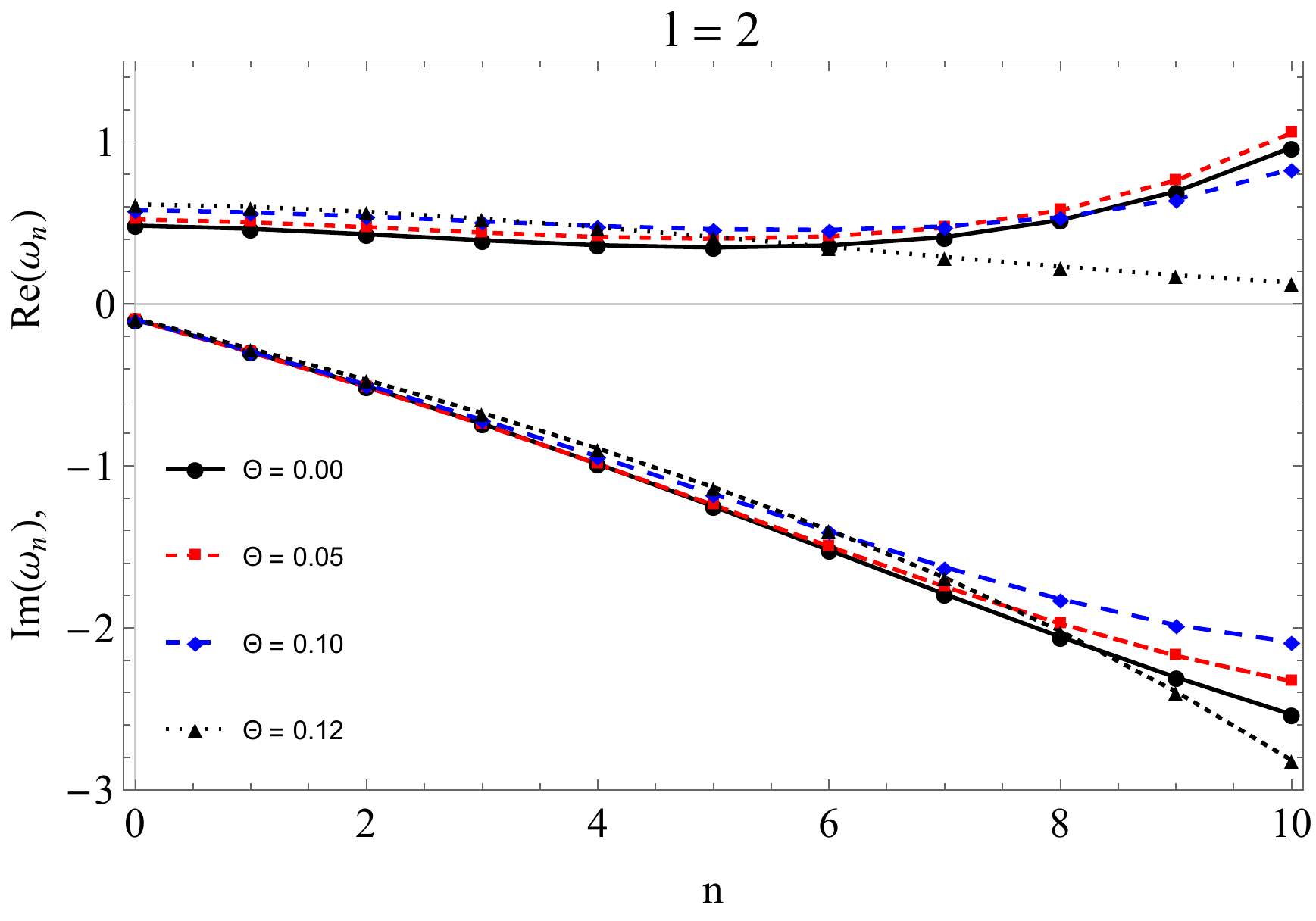}\label{figQN2}}
 \qquad
 \subfigure[]{\includegraphics[scale=0.35]{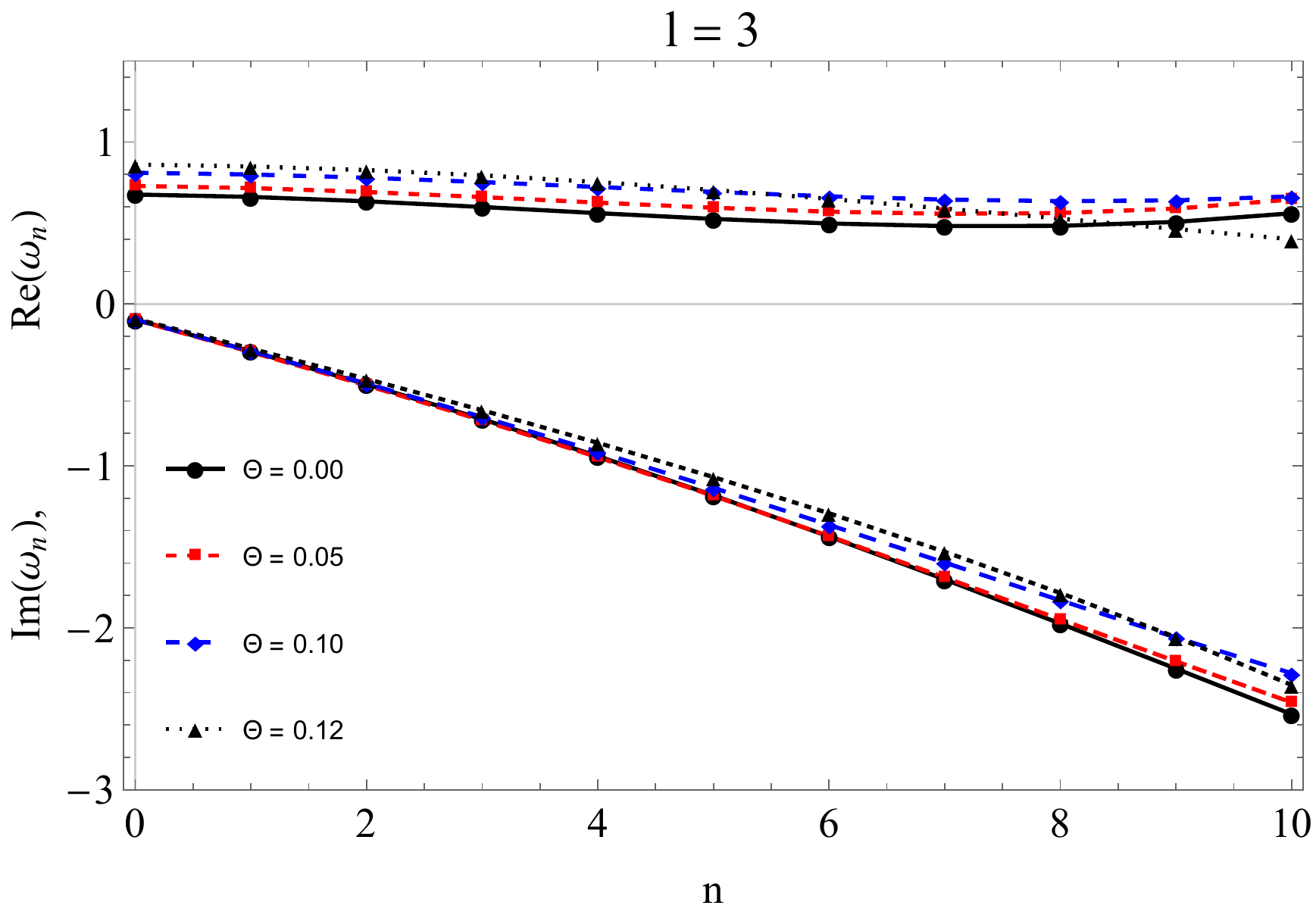}\label{figQN3}}
 \qquad
 \subfigure[]{\includegraphics[scale=0.35]{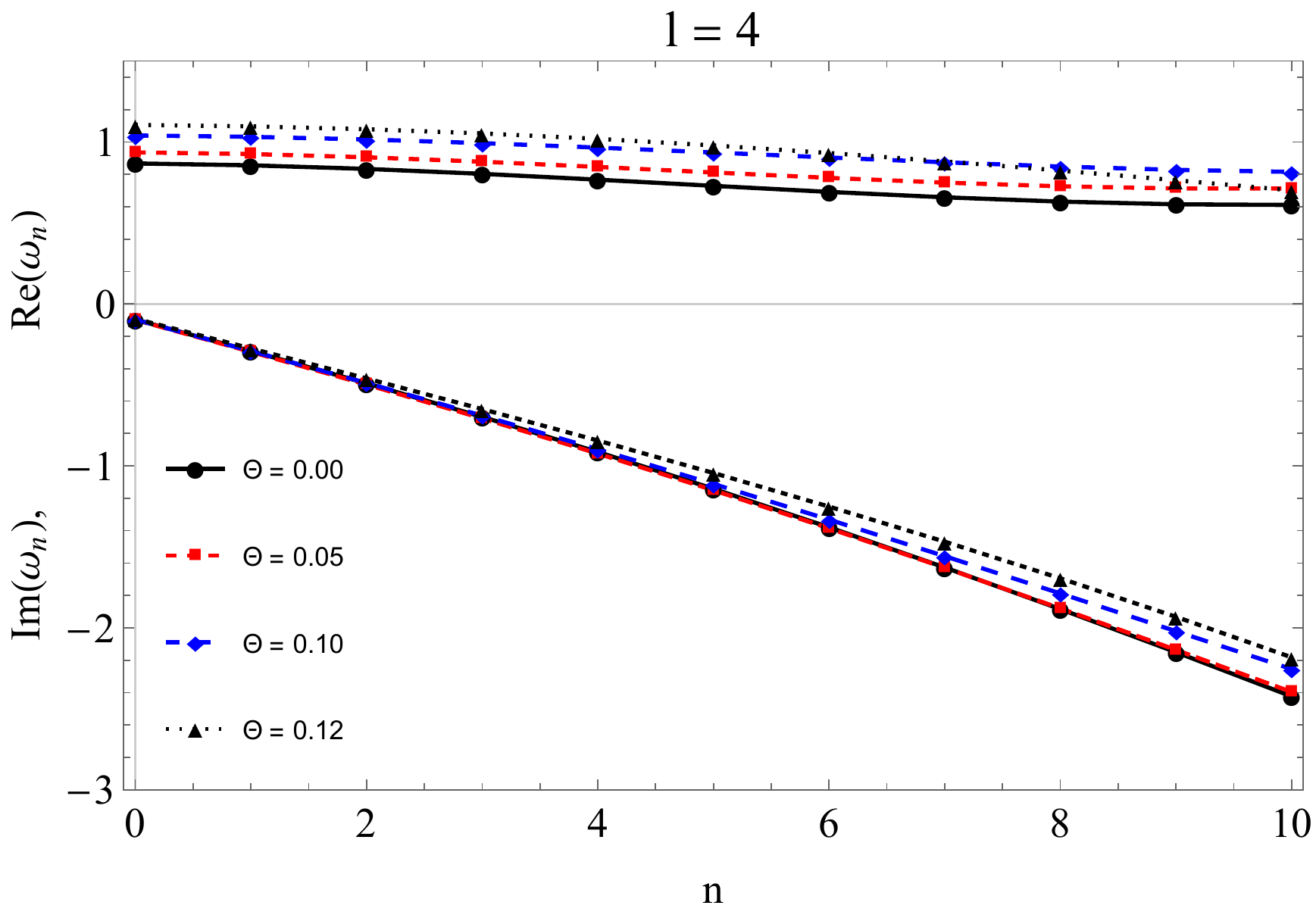}\label{figQN4}}
 \\
 \caption{\footnotesize{Real (top) and imaginary (bottom) parts of the quasinormal frequencias as function of the $n$. We see that for $ l = 1 $ the frequency curves are very dispersed with the increase of $ n $, while for $ l = 3, 4 $ the curves of the real part of the frequency are more constant.}} 
 \label{fig_Qn}
\end{figure}
\begin{figure}[!htb]
 \centering
 \subfigure[]{\includegraphics[scale=0.42]{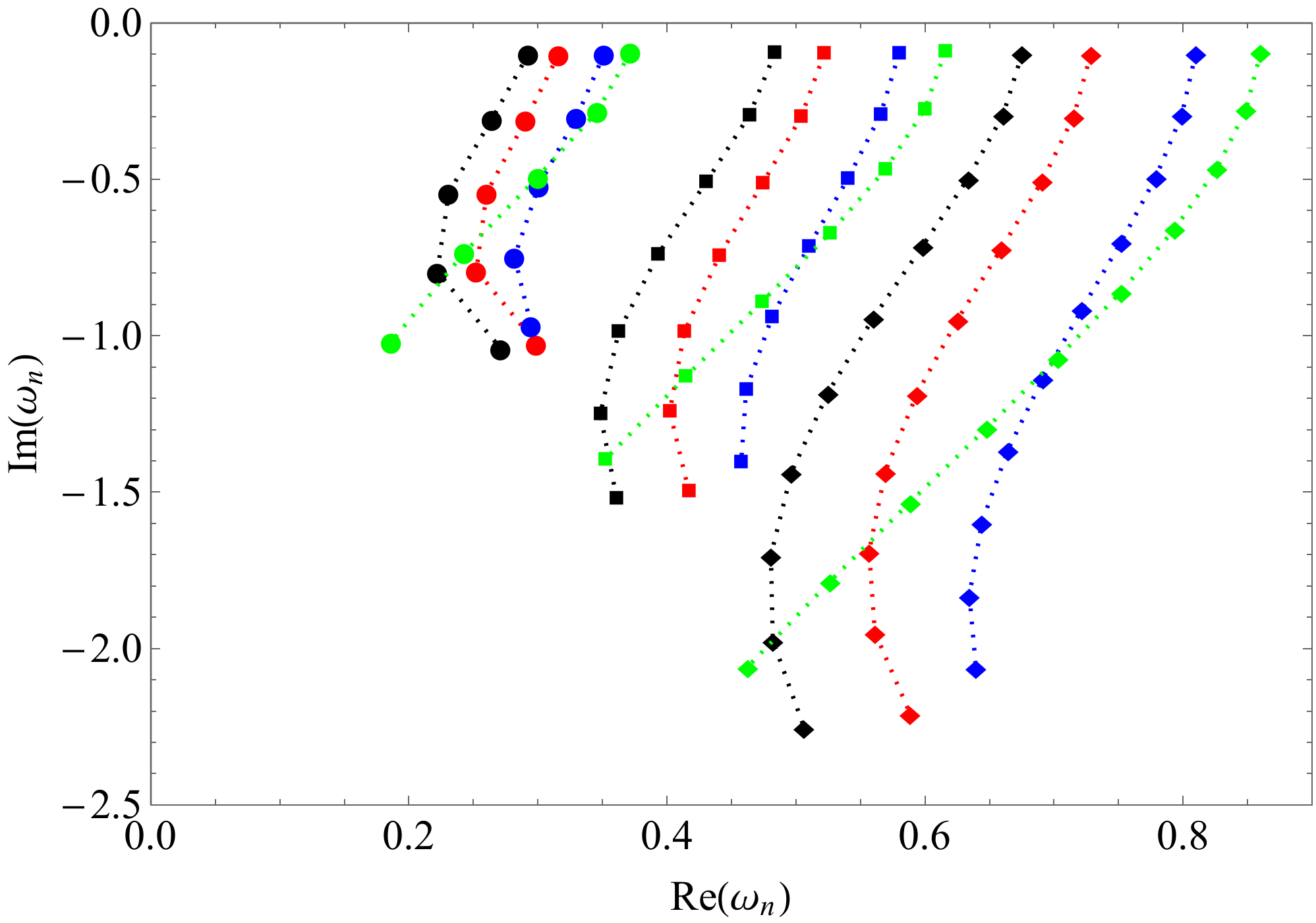}\label{plano}}
 \qquad
 \subfigure[]{\includegraphics[scale=0.42]{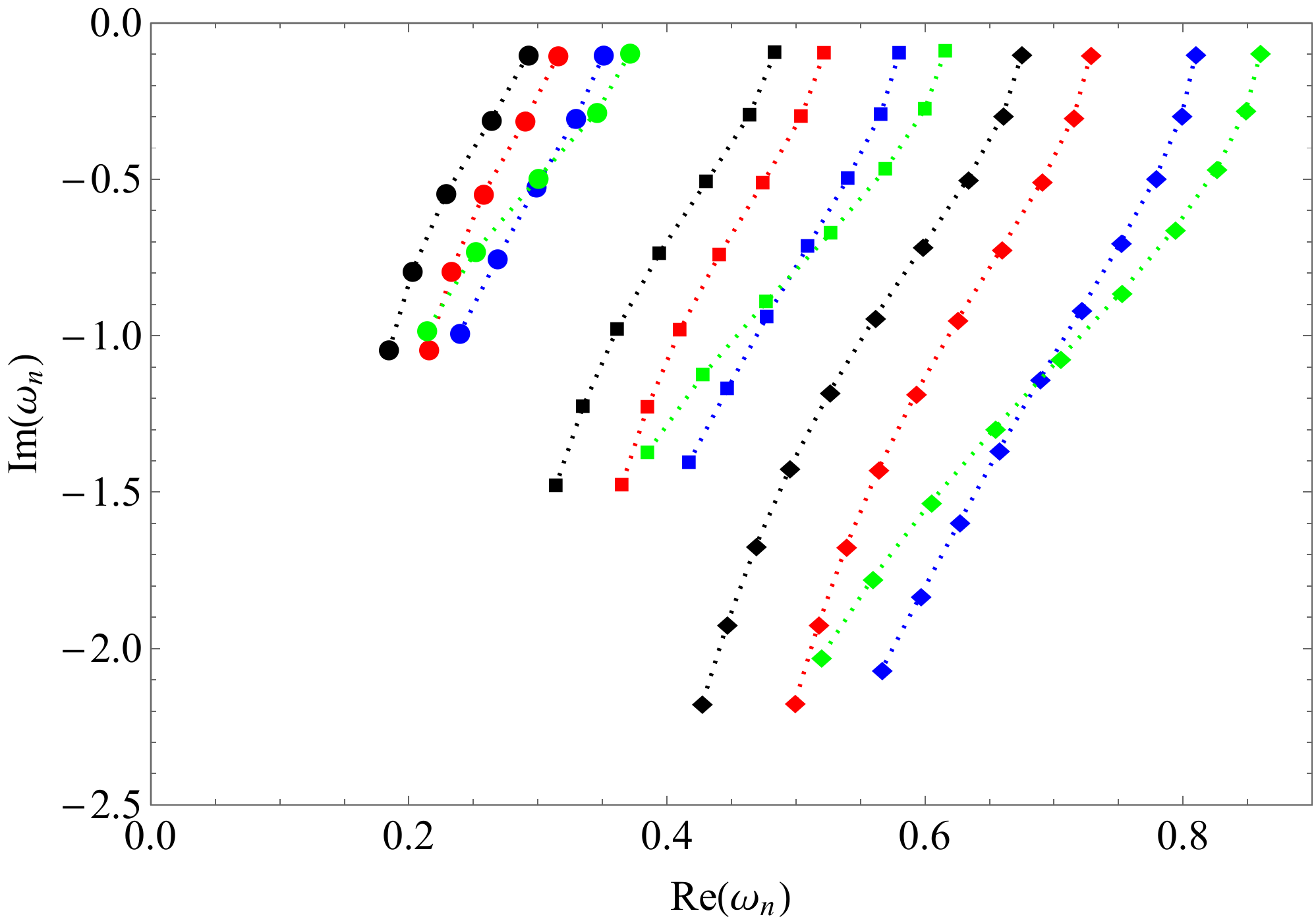}\label{planoNum}}
 \caption{\footnotesize{Complex plane of the QNMs. In plot (a) we have the results obtained by the WKB approximation, while in (b) we use continuous fraction. The markers denote the multipole number as: $l=1$(circle), $l=2$(square) and $l=3$(diamond), while the colors denote the value of the noncomutative parameter $\Theta=0$(black), $\Theta=0.05$(red), $\Theta=0.10$(blue) and $\Theta=0.12$(green).}} 
  \label{plaxoplx}
\end{figure}

\section{Null geodesic and Shadow of a noncommutative black hole}\label{sec3}
We know that in the vicinity of a black hole all the photons are absorbed so that a distant observer looking at the black hole, in absence of any other source,  will see a spot created by this absorption, and this spot is usually called the shadow of the black hole. These shadows have been studied long ago by Synge~\cite{Synge:1966okc} and Luminet~\cite{Luminet:1979nyg} who started studies for Schwarzschild black hole while Kerr black holes were studied by Bardeen~\cite{Bardeen}.

As these shadows correspond to the apparent shape of the photon capture orbits, the space-time metric itself is enough to determine them and thereby better understand the geometry of the near horizon. One method of determining the apparent shape of the black hole is through the shadow boundary that can be studied by the equations of null geodesics. Similar studies have been done in different contexts~\cite{Wei:2015dua,Atamurotov:2015xfa,Shaikh:2019fpu,Tsupko,Stuchlik:2019uvf,Jusufi:2019ltj}

\subsection{Null geodesic}
We can find the geodesics from equation (\ref{elbg}) by taking a Lagrangian in the form
\begin{equation}
\mathcal{L} \equiv \dfrac{1}{2}g_{\mu\nu}\dot{x}^{\mu}\dot{x}^{\nu}.
\end{equation}
Thus, we have
\begin{equation}
2\mathcal{L} = f(r)\dot{t}^{2} - \dfrac{\dot{r}^{2}}{f(r)} - r^{2}\left( \dot{\vartheta}^{2} + \sin^{2}\vartheta \dot{\phi}^{2} \right),
\label{elidot}
\end{equation} 
where the ``$\cdot $" is the derivative with respect to an affine parameter.

We are interested in the path of a ray of light in the described metric, which is spherically symmetrical, so if we analyze in a plane, any ray of light that begins with a certain angle $ \vartheta $ must remain with the same angle. We will then consider an equatorial plane by setting the angle $ \vartheta $ to $ \pi/2 $.

Thus, two equations are enough to describe the movement of a beam of light. 
We can put together a system with these equations that give rise to two geodesic motion constants $ E $ and $ L $, which correspond to energy and angular momentum respectively:
\begin{equation}
E = f(r)\dot{t}, \qquad \quad L = r^{2}\dot{\phi}.
\label{EL}
\end{equation}
Now, as our goal is to study the null geodesics we have to $g_{\mu\nu}\dot{x}^{\mu}\dot{x}^{\nu} = 0$, and using the equations \eqref {EL} we can write
\begin{equation}
\dot{r}^{2}  + f(r)\dfrac{L^{2}}{r^{2}} = E^{2}.
\label{eqEner}
\end{equation}
Introducing a new variable $ u = 1/r $ we can write the orbit equation as follows
\begin{equation}
\dfrac{du}{d\phi} =\sqrt{\dfrac{1}{b^2} - u^{2} + 2M u^{3} - \dfrac{8M\sqrt{\theta}}{\sqrt{\pi}}u^{4}},
\label{eqD1}
\end{equation}
where $ b = L/E $ is the impact parameter defined as the perpendicular distance (measured at infinity) 
between the geodesic and a parallel line that passes through the origin.
So differentiating \eqref{eqD1} we have,
\begin{equation}
\dfrac{d^{2}u}{d\phi^{2}} = - u + 3M u^{2} - \dfrac{16M\sqrt{\theta}}{\sqrt{\pi}}u^{3}.
\label{eqD2}
\end{equation}
By solving the equations \eqref{eqD1} and \eqref{eqD2} numerically, we can obtain the behavior of the geodesic lines for different values of the impact parameter $ b $. 
In the figure \ref{geod}, we verify the change of the geodesic lines for different impact parameters $ b $ 
and also by varying the values of the noncommutative parameter. 
In the figures, we have a black disk that represents the limit of the event horizon, the internal dotted circle is the radius for the photon sphere (critical radius), and the external dashed circle is the critical impact parameter (shadow). 
Hence, we see that the noncommutative parameter decreases the effect of the black hole on the light beams. 
For a similar effect see also~\cite{Crispino:2009ki}. 
\begin{figure}[htbh]
\centering
\subfigure[]{\includegraphics[scale=0.30]{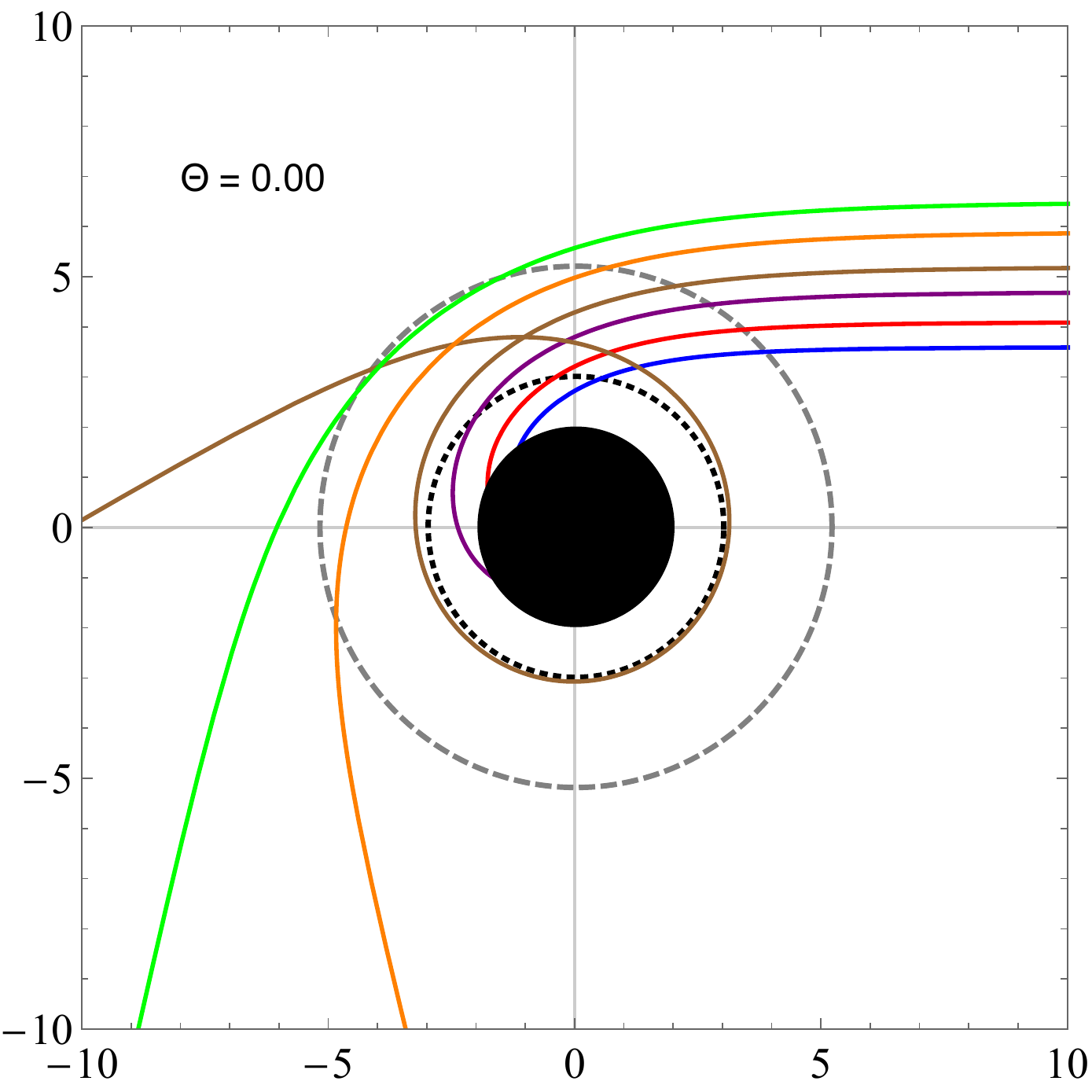}}
\subfigure[]{\includegraphics[scale=0.30]{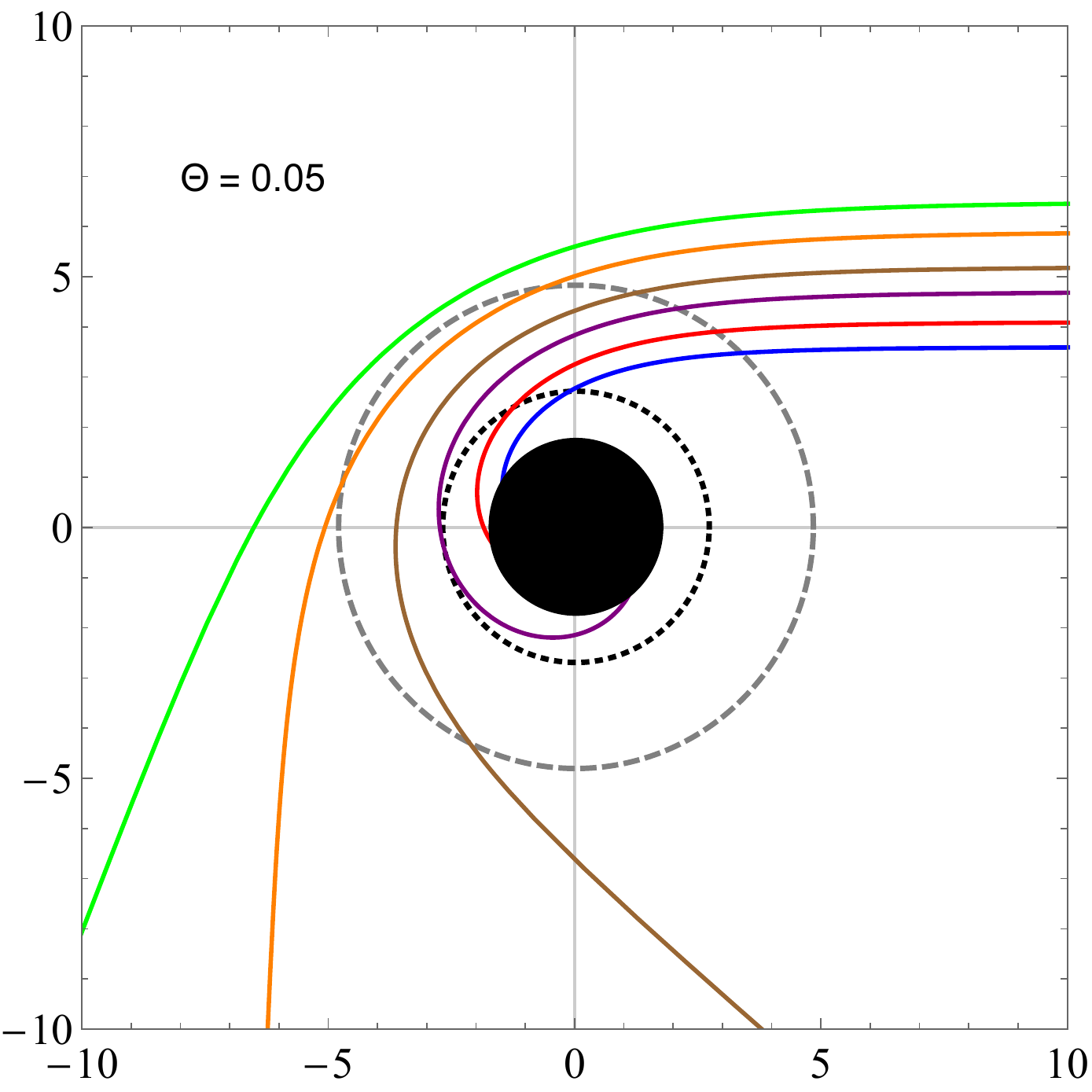}}
\subfigure[]{\includegraphics[scale=0.30]{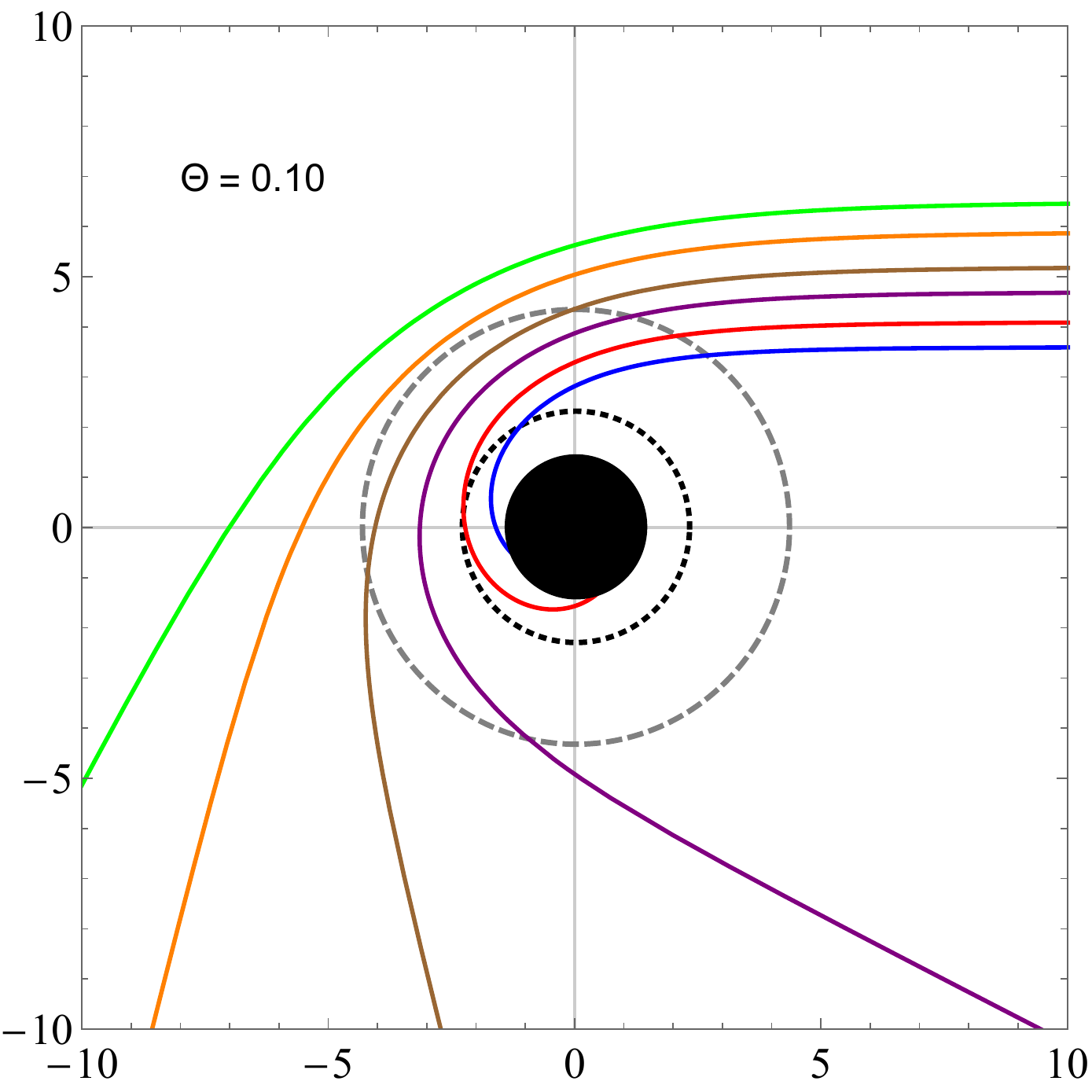}}
\subfigure[]{\includegraphics[scale=0.30]{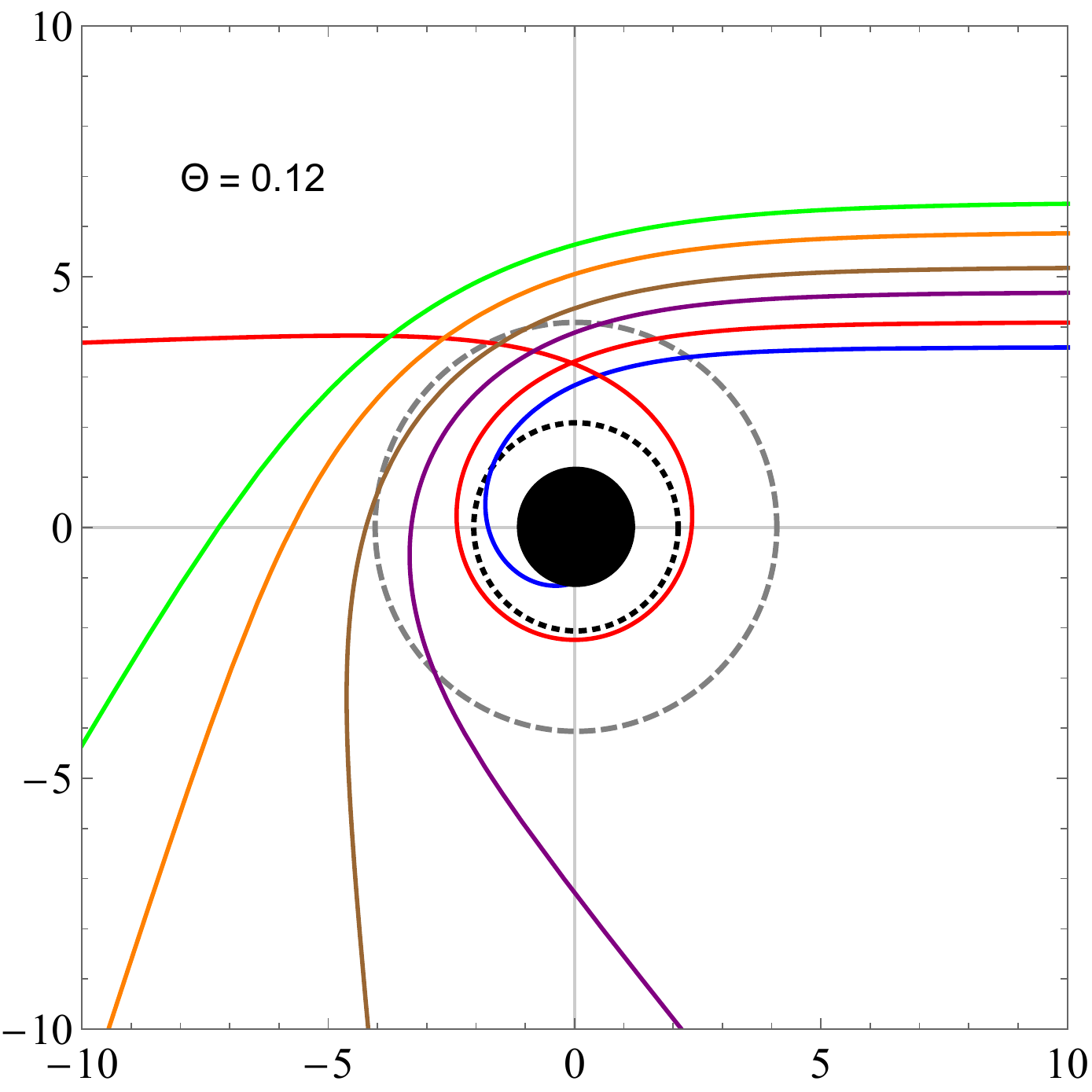}}
\caption{\footnotesize{Geodesics surrounding a noncommutative black hole. 
The impact parameters defined as $b = 3.6, 4.1, 4.7, 5.2, 5.9$ and $6.5$ are the same for all graphs, assuming $M = 1$. We can clearly see the influence of the noncommutative parameter on the geodesic curves from 
(a) $\Theta = 0$ (Schwarzschild case) to (d) $\Theta = 0.12$. }}
 \label{geod}
\end{figure}   
\subsection{Critical orbit and shadows}
It is known that the shadow of the black hole is directly related to the impact parameter for the photon orbit, as we will see below. So to determine the shadow limit, we will start by studying the effective potential that satisfies the equation of the null geodesic as follows
\begin{equation}
\dot{r}^{2} +  {\cal V}_{eff}(r) = 0,
\label{eqVnull}
\end{equation}
where using \eqref{eqEner}, we have 
\begin{eqnarray}
{\cal V}_{eff}(r) = f(r)\dfrac{L^{2}}{r^{2}} - E^{2}.
\end{eqnarray}
In this case, we can obtain a critical radius or critical circular orbit for a photon $ r_{c} $ and critical impact parameter $ b_{c} $, by using the following conditions: 
${\cal V}_{eff}(r_{c}) = 0$ and $\dfrac{d{\cal V}_{eff}(r_{c})}{dr} = 0$. 
{ So we find
\begin{eqnarray}
r_{c} &=& \frac{3}{4}\left(r_{+} + r_{-}\right) +
\frac{3}{4}\sqrt{(r_{+} + r_{-})^2 - \frac{32r_{+}r_{-}}{9}},
\\
&=&\frac{3M}{2}+\frac{3M}{2}\sqrt{1-\frac{64\sqrt{\theta}}{9M\sqrt{\pi}}},
\\
b_{c} &=& \dfrac{r_{c}}{\sqrt{f(r_{c})}} = \frac{r^2_c}{\sqrt{(r_c - r_+)(r_c-r_-)}}.
\label{paramcrit}
\end{eqnarray} 
Let us now compute  the size of the black hole shadow that can be expressed via celestial coordinates as follows~\cite{Bardeen}
\begin{eqnarray}
\alpha &=& \lim\limits_{r_{o} \to \infty}\left[- r_{o}^{2} \sin\vartheta_{o}\dfrac{d\phi}{dr}\Bigr\rvert_{\vartheta = \vartheta_{o}} \right],\\
\beta &=& \lim\limits_{r_{o} \to \infty}\left[r_{o}^{2} \dfrac{d\vartheta}{dr}\Bigr\rvert_{\vartheta = \vartheta_{o}}\right],
\label{coodcelest}
\end{eqnarray}
where $\left(r_{o}, \vartheta_{o}\right)$ is the observer position at infinity. 

As our study is restricted to the equatorial plane, the radius that delimits the size of the shadow is equivalent to the critical impact parameter, and so we have
\begin{equation}
R_{s} \equiv \sqrt{\alpha^{2} + \beta^{2}} = b_{c},
\end{equation}
where
\begin{equation}
\label{rs}
R_s= \left[\dfrac{\left( 3M + \sqrt{9M^2 -{64M\sqrt{\theta}}/{\sqrt{\pi}} + {128{\theta}}/{{\pi}}} \right)^4}
{8\left( 3M^2 -{16M\sqrt{\theta}}/{\sqrt{\pi}}+{32 \theta}/{\pi} + M\sqrt{9M^2 -{64M\sqrt{\theta}}/{\sqrt{\pi}} + {128{\theta}}/{{\pi}}} \right)}\right]^{1/2}.
\end{equation}

{In a semiclassical description of the scattering~\cite{Ford1959}, the impact parameter is associated with each partial wave $b = (l + 1/2)/\omega$ in the large $l$ regime.
{As shown in~\cite{cardoso2009geodesic},  the real part of the quasinormal frequencies at the eikonal limit 
corresponds to the angular velocity for the last null circular orbit $ \Omega_c $ and the imaginary part is associated with the Lyapunov exponent $ \lambda $ which determines the unstable timescale of the orbit
\begin{eqnarray}
\omega_{QNM}=\Omega_c l - i\left(n +\frac{1}{2}\right)|\lambda|,
\end{eqnarray}
being the angular velocity given by }
\begin{eqnarray}
\Omega_c=\frac{\dot{\phi}}{\dot{t}}=\frac{f(r_c)b_c}{r^2_c}=\frac{1}{b_c}
\end{eqnarray}

This suggests a relationship between quasinormal frequencies and shadow radius as done by Jusufi~\cite{jusufi2020quasinormal} at the eikonal limit, showing that the real part of quasinormal modes is inversely proportional to the radius $R_{s}$ as follows 
\begin{equation}
Re(\omega) = \lim_{l>>1}\dfrac{l}{R_{s}}.
\label{eq_eikonal}
\end{equation}
Note that this expression is valid only for large values of $l$ in most of the cases although fails for Einstein-Lovelock
theory as shown by Konoplya and Stuchlik~\cite{Konoplya:2017wot}. We can see in Fig.~\ref{fig_Qn} that the higher the value of $l$ the smaller the contribution of $n$. Thus by using the results obtained by the WKB approximation we compare with the shadow radius $R_{s}$.}
\begin{table}[h!]
\begin{center}
\begin{footnotesize}
\caption{\footnotesize{The real part of quasinormal frequencies for large $l$ compared with shadow radius $R_{s}$ }}
\label{tab4}
\begin{tabular}{ c||c|c|c  }
 \hline
\multicolumn{1} {c||}{$M = 1$}&\multicolumn{2} {|c|}{ $Re(\omega)/\left(l+1/2\right)$ } & \multicolumn{1} {|c}{ $\left( R_{s}\right)^{-1}$ } \\
\hline
$\Theta$   &  $l = 100$ & $l = 1500$  &  $----$ \\
 \hline
 0.00  & 0.192450707 & 0.192450092 & 0.192450090  \\
 0.05  & 0.207583654 & 0.207582949 & 0.207582946  \\
 0.10  & 0.230750607 & 0.230749836 & 0.230749832  \\
 0.12  & 0.245314933 & 0.245314347 & 0.245314347  \\
 \hline
 \end{tabular}
 \end{footnotesize}
\end{center}
\end{table}

{In the table \ref{tab4} we see that by increasing the value of $l$, the results between the real part of the quasinormal frequencies and the black hole shadow radius approach each other.} Now, we can express $ R_{s} $ by considering $ \theta $ small, and so we get the following approximate expression
\begin{equation}
 R_{s} \approx 3\sqrt{3}M - 4\sqrt{3}\sqrt{\frac{\theta}{\pi}}
  + \frac{16}{3\sqrt{3}M}\frac{\theta}{\pi} + \cdots . 
\end{equation}
Note that for $ \theta = 0 $, we have the shadow radius for the Schwarzschild black hole case. 
Therefore, we notice that the shadow radius is reduced when we change the parameter $ \theta $.
In Fig.~\ref{sha}, we see the circles that represent the shadow boundaries of the noncommutative black hole for different values of $ \Theta $. 
Note that we have a reduction in the circles when we vary $ \Theta $.
 Furthermore, taking $M \rightarrow 0$ in (\ref{rs}), we obtain a non-zero result for the shadow radius, that is
\begin{equation}
 R_{s} \approx 8M_{min}, 
\end{equation}
where $ M_{min}=\sqrt{\theta/\pi} $ is the minimal mass~\cite{Anacleto:2020zfh}. 
Therefore, at the limit of $ M\rightarrow 0  $ the shadow radius is proportional to the minimum mass and the black hole becomes a black hole remnant.
We have shown this behavior in Fig.~\ref{shadt}.
In Fig.~\ref{shadt}, we show the behavior of the shadow radius by keeping $ \Theta $ fixed and assuming small values of $ M $. 
In Fig.~\ref{shadm}, We show the behavior of the shadow radius by keeping $ M $ fixed and assuming small values of $ \Theta $. 
}
\begin{figure}[htbh]
\centering
\includegraphics[scale=0.45]{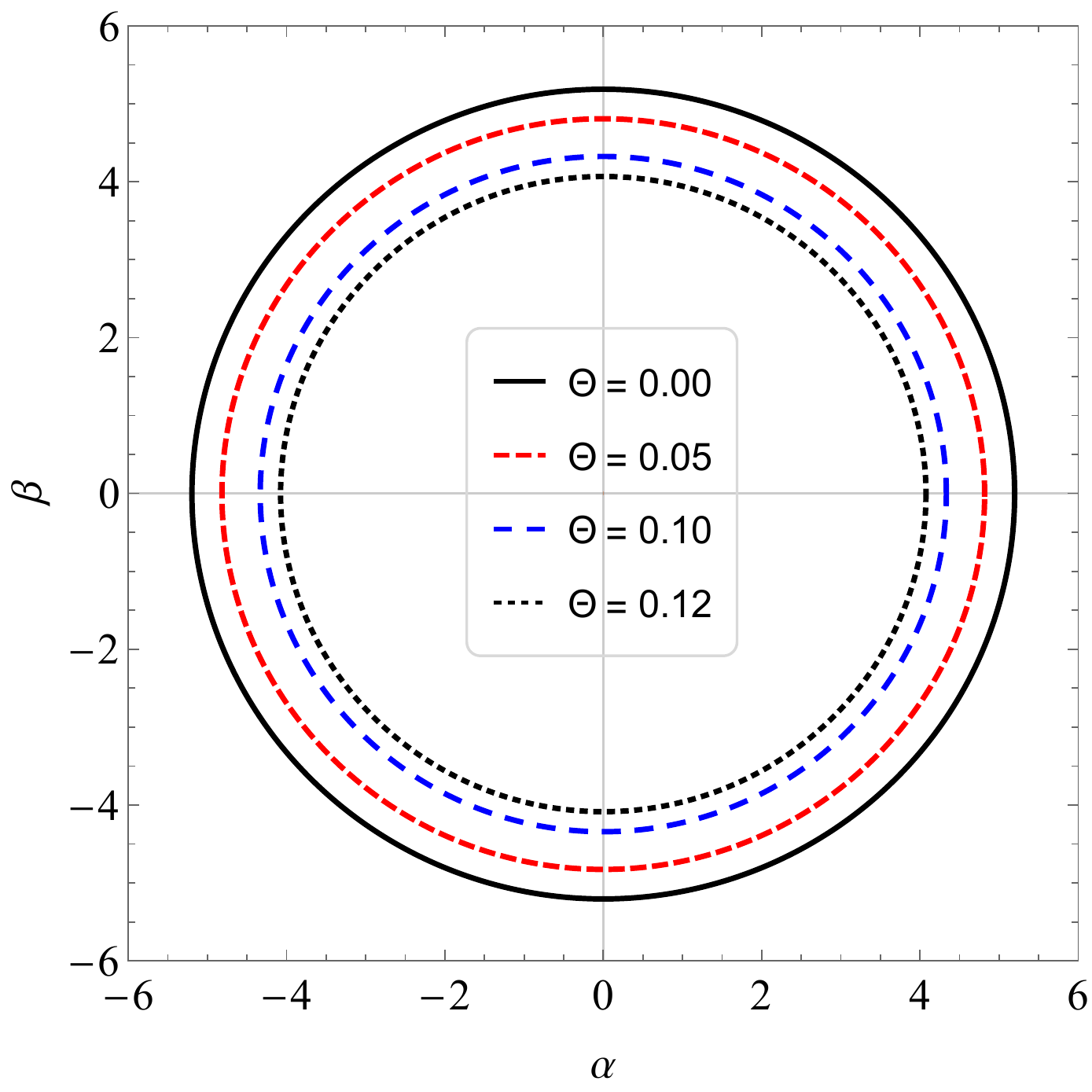}
 \caption{We see the influence of non-commutativity in the shadow admitting 
$ M=1 $ and $\Theta = 0.0, 0.05, 0.10$ and $0.12$. }
 \label{sha}
\end{figure} 
\begin{figure}[!htb]
 \centering
\subfigure[]{\includegraphics[scale=0.4]{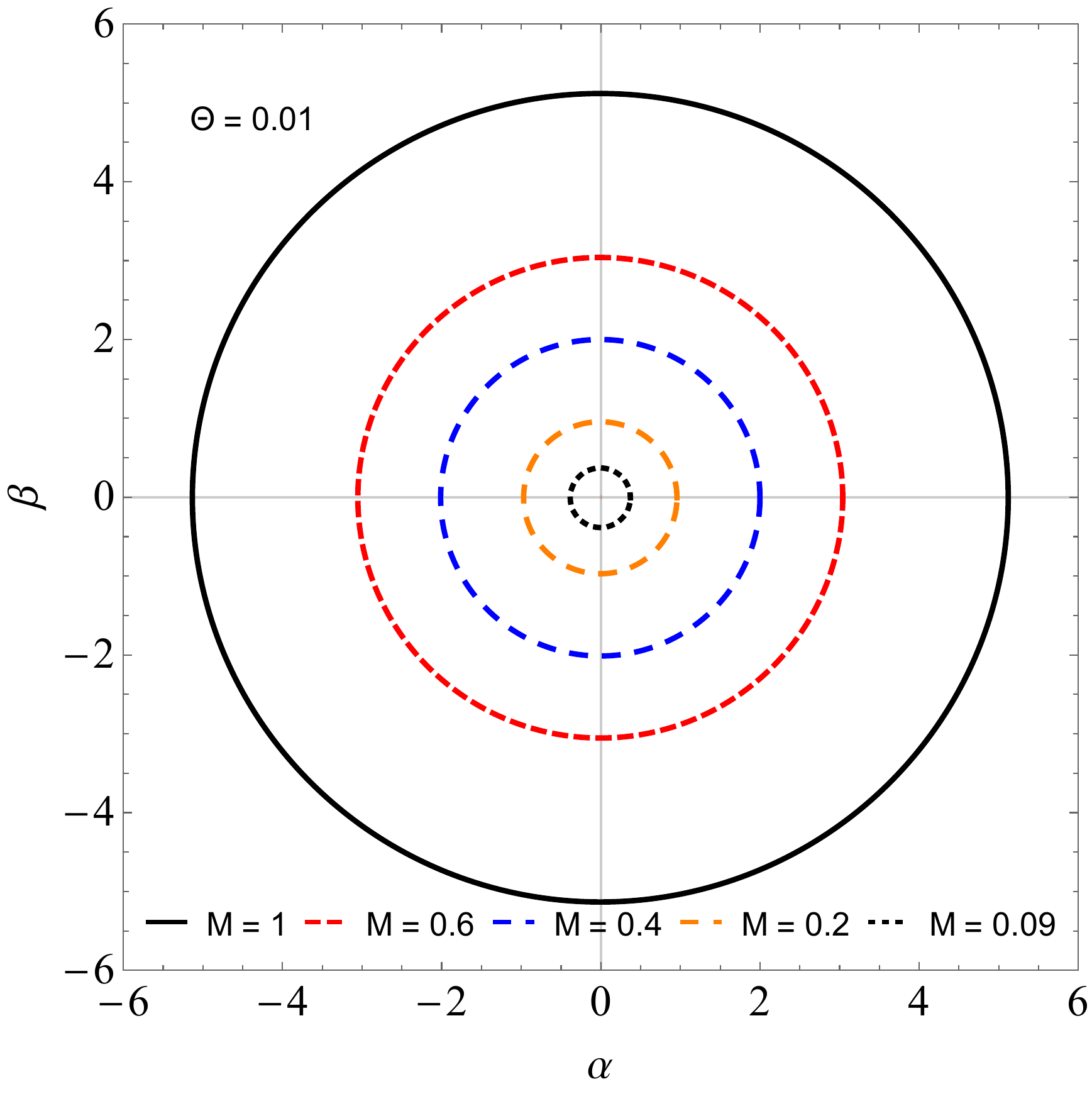}\label{shadta}}
\qquad
\subfigure[]{\includegraphics[scale=0.4]{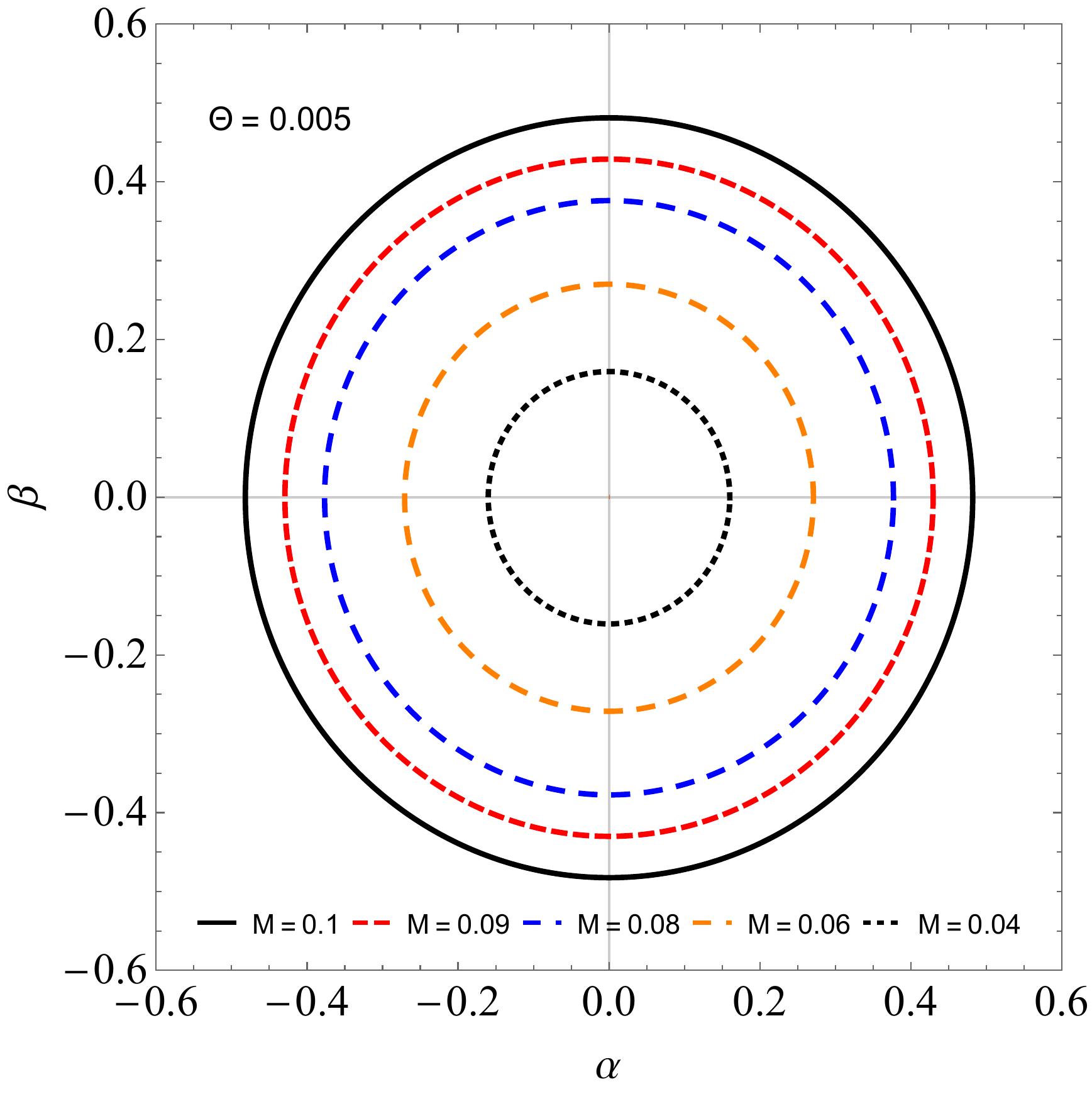}\label{shadt2}}
 \caption{We see the influence of non-commutativity in the shadow admitting 
(a) $\Theta = 0.01 $ and $M=1.0, 0.6, 0.4, 0.2, 0.09 $. 
(b) $\Theta = 0.005 $ and $M=0.1, 0.09, 0.08, 0.06, 0.04 $.} 
 \label{shadt}
\end{figure}
\begin{figure}[!htb]
 \centering
\subfigure[]{\includegraphics[scale=0.4]{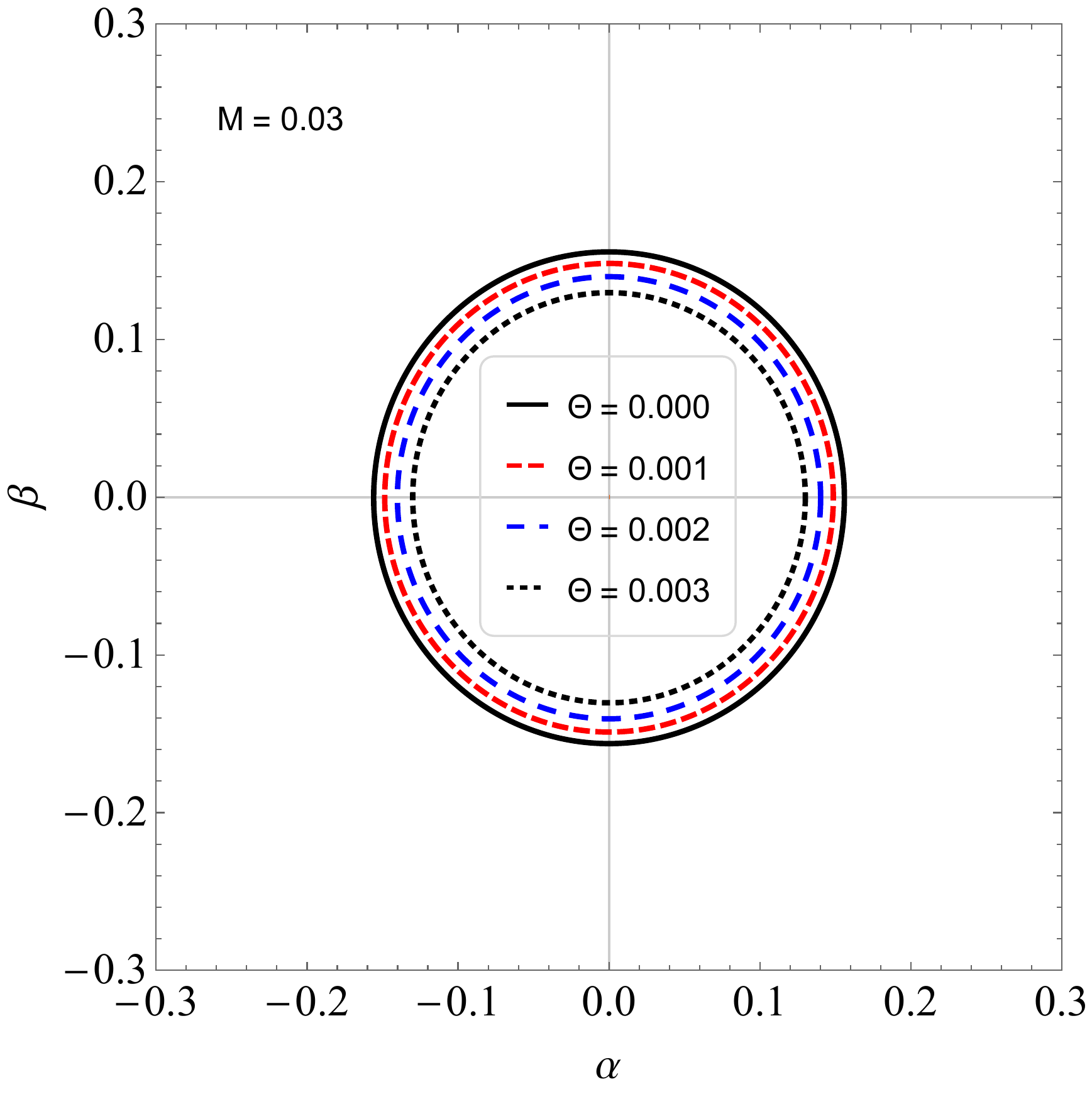}\label{shadma}}
\qquad
\subfigure[]{\includegraphics[scale=0.4]{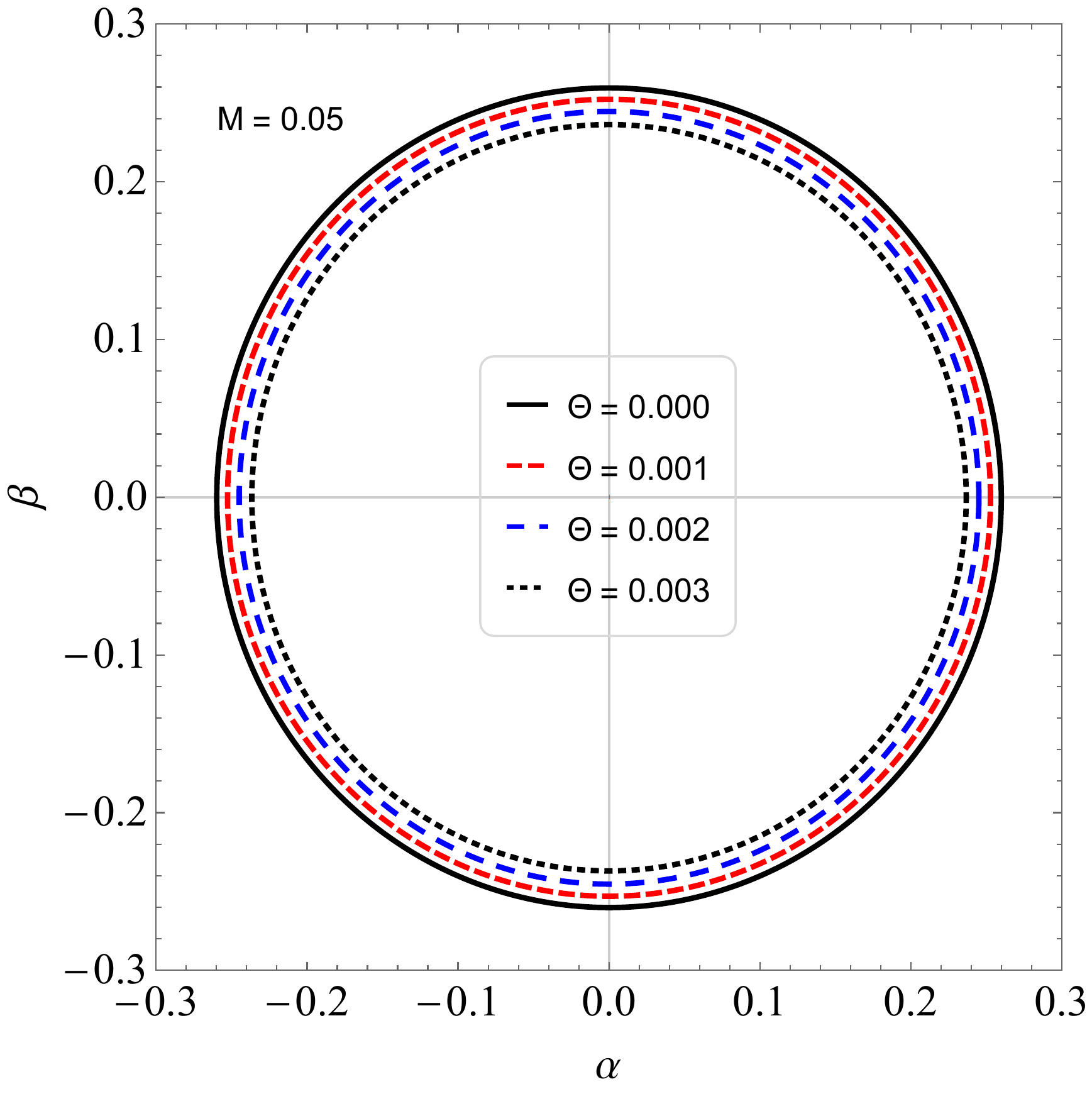}\label{shadm2}}
 \caption{We see the influence of non-commutativity in the shadow admitting 
(a) $M = 0.03 $ and $\Theta=0.000, 0.001, 0.002, 0.003 $. 
(b) $M = 0.05 $ and $\Theta=0.000, 0.001, 0.002, 0.003 $.} 
 \label{shadm}
\end{figure}

\section{Conclusions}\label{conc}
In summary, in this work, we investigate the quasinormal frequencies for a noncommutative Schwarzschild black hole by two different methods in order to investigate and compare the results. Using the sixth-order WKB approximation and Leaver's continuous fraction, we found that there is a small difference between the quasinormal frequencies obtained by each method mainly for small multipoles. The effects of the noncommutative parameter $\Theta$ cause an increase in the real part of the quasinormal frequencies, while the magnitude of the imaginary part begins to grow and then decreases.
{For the black hole shadow we use the results obtained by the WKB method to verify that in large $l$ regimes the real part of quasinormal modes is inversely proportional to the shadow radius.}
{In addition, we have shown that the shadow radius is non-zero at the zero mass limit. 
Therefore being proportional to a minimum mass.
However for $\theta=0$, we recover the shadow radius for the Schwarzschild black hole case. Finally, we also notice that the shadow radius is reduced when we increase the noncommutative parameter.}

\acknowledgments
We would like to thank CNPq, CAPES and CNPq/PRONEX/FAPESQ-PB (Grant nos. 165/2018 and 015/2019),  for partial financial support. MAA, FAB and EP acknowledge support from CNPq (Grant nos. 306962/2018-7 and  433980/2018-4, 312104/2018-9, 304852/2017-1).

\section{Appendix}\label{Append}
Here we will show that the differential equation \eqref{ER} can be written in the form of a generalized spheroidal wave equation, as done by Leaver for the Schwarzschild and Kerr cases in~\cite{leaver1986solutions}. The generalized spheroidal wave equation has the following form
\begin{equation}
x(x-x_{0})\dfrac{d^{2}y}{dx^{2}} +\left(B_{1} + B_{2}x \right)\dfrac{dy}{dx} + \left[\omega^{2} x(x-x_{0}) - 2\eta\omega(x - x_{0}) + B_{3}\right]y = 0.
\label{EReph}
\end{equation}
The radial equation \eqref{ER} can be rewritten as follows
\begin{eqnarray}
 \left(r - r_{-}\right)\left(r - r_{+}\right)\dfrac{d^{2}R(r)}{dr^{2}} + \dfrac{\left(r(r_{+} + r_{-}) - 2r_{-}r_{+}\right)}{r}\dfrac{dR(r)}{dr} + \Big[\dfrac{\omega^{2}r^{4}}{\left(r - r_{-}\right)\left(r - r_{+}\right)} - \dfrac{\left(r(r_{+} + r_{-}) - 2r_{-}r_{+}\right)}{r^{2}}
 \nonumber\\
   - l(l+1)\Big]R(r) = 0.
\label{ERappen}
\end{eqnarray}
In this appendix, by considering $2M= 1$, for simplicity, we have $r_{\pm} = \left(1 \pm b\right)/2$, where $b = \sqrt{1 - 16\sqrt{\theta /\pi}}$. Firstly, we have to find a suitable transformation, by using the following solution
\begin{eqnarray}
R(r) = r(r-r_{-})^{\alpha}(r -r_{+})^{\beta}y(r-r_{-}),
\label{solR}
\end{eqnarray}
where $\alpha$ and $\beta$ are parameters to be determined. Furthermore, let us assume that $x = r - r_{-}$, with $x \rightarrow 0 $ as $r\rightarrow r_{-}$ and $x \rightarrow b $ as $r\rightarrow r_{+}$. Now applying these changes we have that the equation \eqref{ERappen} can be written in the form
\begin{equation}
 x(x-b)\dfrac{d^{2}y}{dx^{2}} +\left[b(1+\alpha) + 2(1+ \alpha + \beta)x \right]
 \dfrac{dy}{dx} + U(x)y = 0,
\label{ERappen2}
\end{equation} 
where 
\begin{eqnarray}
U(x) = &-&\dfrac{\left[(b-1)^{4}\omega^{2}+16b^{2}\alpha^{2}\right]}{16bx} + \dfrac{\left[(b+1)^{4}\omega^{2}+16b^{2}\beta^{2}\right]x}{16 b^{2}\left(x - b\right)} + \dfrac{\left(b-1\right)^{3}\left(1+7b\right)\omega^{2}}{16b^{2}} +(2-b)x\omega^{2} + x^{2}\omega^{2}\nonumber \\ 
&+& \alpha +2\alpha^{2} + \beta + 2\alpha- l(l+1).
\label{solU}
\end{eqnarray}
In order to have an equation of the form \eqref{EReph} the first two terms of \eqref{solU} must be zero, that is, $(b-1)^{4}\omega^{2} + 16b^{2}\alpha^{2} = 0$ and $(1+b)^{4}\omega^{2} + 16b^{2}\beta^{2} = 0$, so we have the following values for $\alpha $ and $\beta$
\begin{eqnarray}
\alpha &=& -\dfrac{i(b-1)^{2}\omega}{4b}, \qquad \dfrac{i(b-1)^{2}\omega}{4b}\\
\beta &=& -\dfrac{i(b+1)^{2}\omega}{4b}, \qquad \dfrac{i(b+1)^{2}\omega}{4b}.
\end{eqnarray}
Thus, for $\alpha = -i(b-1)^{2}\omega/4b = -ir_{-}^{2}\omega/b$ and $\beta = -i(b+1)^{2}\omega/4b = -ir_{+}^{2}\omega/b$, the equation \eqref{ERappen2} is of the form
\begin{eqnarray}
\label{ERappen3}
 x(x-b)\dfrac{d^{2}y}{dx^{2}} + \left[-b+2ir_{-}^{2}\omega + \dfrac{2}{b}(b - i\omega(r_{-}^{2}+r_{+}^{2}))x \right]
 \dfrac{dy}{dx} + \left[\omega^{2}(x-b)x + 2\omega^{2}(x-b) \right.\\ 
  \left. + \dfrac{\omega^{2}}{4b}\left(-1+4b+4b^{2}+b^{4}\right)\nonumber -\dfrac{i\omega}{2b}\left(b^{2}+1\right)- l(l+1)\right]y = 0.
\end{eqnarray}
This implies that the equation \eqref{ERappen} can be transformed into a generalized spheroidal equation. By comparing it with \eqref{EReph} we have:
\begin{eqnarray}
B_{1} = -b+2ir_{-}^{2}\omega, \qquad B_{2}= \dfrac{2}{b}(b - i\omega(r_{-}^{2}+r_{+}^{2})),\\
B_{3}= \dfrac{\omega^{2}}{4b^{2}}\left(-1+4b^{2}+4b^{3}+b^{4}\right)-\dfrac{i\omega}{2b}\left(b^{2}+1\right)- l(l+1)\qquad \text{and} \qquad \eta = -\omega.
\end{eqnarray}
Thus, by using Jeffe's regular solution to $y(x)$ as in \cite{leaver1986solutions} (p.~12) for a generalized spheroidal wave equation we have
\begin{eqnarray}
y(x) = e^{i\omega x}x^{-1+i\omega(r_{-}^{2}+r_{-}^{2})/b + i\omega}\sum_{k=0}^{\infty}a_{k}\left(\frac{x - b}{x}\right)^{k}.
\label{solyx}
\end{eqnarray} 
We can also compare the solution \eqref{solR} with the values of $\alpha$ and $\beta$ to the solution \eqref{solyx}, by rewriting it in the form
\begin{equation}
R(r) = \frac{r}{r - r_{-}}\left(r - r_{-}\right)^{i \omega}\left(\frac{r - r_{+}}{r - r_{-}}\right)^{\frac{-i \omega r_{+}^{2}}{b}} e^{i\omega (r-r_{-})}\sum_{k=0}^{\infty}a_{k}\left(\frac{r - r_{+}}{r - r_{-}}\right)^{k}.
\label{newR}
\end{equation}
This is precisely our solution \eqref{anzart}, for $2M=1$. Thus,  just as we have three coefficients in equation (41) in \cite{leaver1986solutions}, we can also get the three coefficients for the recurrence relation of the equation \eqref{ERappen3}. So by using the solution \eqref{solyx} we have
\begin{eqnarray}
\alpha_{0}a_{1} + \beta_{0}a_{0} = 0, \qquad \alpha_{k}a_{k+1} + \beta_{k}a_{k} + \gamma_{k}a_{k-1} = 0, \qquad k \geq 1,
\end{eqnarray}
where
\begin{eqnarray}
\alpha_{k} &=& (1 + k)\left[b(1 + k)- 2ir_{+}^{2}\omega\right]/b, \\
\beta_{k} &=& - r_{+}\left[1 + l(l + 1) + 2k(k + 1) - 8r_{+}^{2}-4i\omega r_{+} -8i\omega k r_{+} \right]/b \\
& & + r_{-}\left[ 1 + l(l+1) + 2k(k + 1) - 2i\omega r_{+} - 4i\omega n r_{+}\right]/b ,\nonumber\\
\gamma_{k} &=& (k - 2i\omega)\left(bk - 2i\omega r_{+}^{2} \right)/b.
\label{coef}
\end{eqnarray}
These coefficients are the same as those found in \eqref{cofrel1}, \eqref{cofrel2} and \eqref{cofrel3}, for $2M=1$.

\end{document}